\documentclass[aps,pre,twocolumn,groupedaddress,showpacs,floatfix]{revtex4}
\usepackage{hyperref}
\usepackage{graphicx}
\usepackage{amsmath}
\usepackage{amsfonts}
\DeclareMathSymbol{\Reals}{\mathbin}{AMSb}{"52}

\renewcommand{\vec}[1]{\mbox{\boldmath{$ #1 $}}}
\newcommand{\ten}[1]{\mbox{\boldmath{$ #1 $}}}
\begin{document}

\title{Projection-operator formalism and coarse-graining}
\author{E.A.J.F.\ Peters}
\email{e.a.j.f.peters@tue.nl}
\affiliation{Dept.\ of Chemical Engineering\\
Technische Universiteit Eindhoven\\
P.O. Box 513\\
5600 MB Eindhoven\\
The Netherlands}
\date{\today}

\begin{abstract}
A careful derivation of the generalized Langevin equation using ``Zwanzig flavor'' projection operator formalism is presented.
We provide arguments why this formalism has better properties compared to alternative projection-operator formalisms for deriving non-equilibrium, non-thermodynamic-limit, equations.
The two main ingredients in the derivation are Liouville's theorem and optimal prediction theory.

As a result we find that equations for non-equilibrium thermodynamics are dictated by the formalism once the choice of coarse-grained variables is made.
This includes a microcanonical entropy definition dependent on the coarse-grained variables.
Based on this framework we provide a methodology for succesive coarse-graining.
As two special cases, the case of linear coefficients and coarse-graining in the thermodynamic limit are treated in detail.
In the linear limit the formulas found are equivalent with those of homogenization theory.

In this framework there are no restrictions with respect to the thermodynamic-limit or nearness to equilibrium.
We believe the presented approach is very suitable for the development of computational methods by means of coarse-graining from a more detailed level of description.
\end{abstract}


\maketitle

\section{Introduction}

This work started out as a investigation into the microscopic basis of the GENERIC formalism \cite{Ott97_2,Grm97,Ott05}.
The GENERIC formalism is usually presented as a generic formulation of non-equilibrium thermodynamics.
The book \cite{Ott05} gives a refinement of the formalism compared to the earlier publications \cite{Grm97,Ott97_2}, especially for the case where fluctuations are important.
Already several publications \cite{Ott98, Pab01, Ott05} have made a link between GENERIC building blocks and microscopic expressions.
However, at several points these derivations make not fully justified approximations.

Instead of taking the formalism as given and try to collect evidence for its correctness we take a different route.
Here we use a constructive approach.
From bottom up, we construct a non-equilibrium thermodynamics theory.
By performing this exercise we find many results that are in itself interesting.The final result is a set of equations that are close to the equations presented by H.C. \"{O}ttinger \cite{Ott05}.
We find that the GENERIC formalism presents a special, important, case.

The GENERIC formalism tries to embed many of the non-equilibrium approaches that can be found in literature.
It does not make a choice on, e.g., the use of microcanonical or canonical ensembles.
It requires that the final equations has a certain structure, e.g., the Poisson-structure of the reversible part of the dynamics.
Some of these requirements come from the philosophy that the coarse-grained equations should inherit as much as possible structure/symmetry from the underlying microscopic equations of motion.

In our constructive approach we find, similarly to some of the approaches to equilibrium thermodynamics, that the microcanonical ensemble should be taken as a starting point.
We find that in non near-to-equilibrium cases the Zwanzig projection operator method (based on the microcanonical ensemble) allows for better approximations than projection-operator methods which we label as Mori, Robertson and Grabert.
In this sense our approach is more restrictive than the GENERIC formalism.
Once a set of coarse-variables is chosen the full framework is fixed.
What we are not concerned with here is the question what constitutes an optimal choice of coarse-grained variables.

Most of the GENERIC framework, such as degeneracy conditions follow from the derivations irrespective of the choice of variables.
One feature we could not proof for the general case is the ``Poisson-structure'' of the reversible part of the dynamics.
In practice, however, it turns out that continuum equations such as hydrodynamic equations have this structure.
Obviously it would be very nice if the Poisson-structure survives coarse-graining always.
The claim of the GENERIC formalism seems to be that one is always able to choose a set of suitable variables such that this structure does survive.

To our opinion, it remains to be proved that the Poisson-structure can be retained for intermediate, mesoscopic scale coarse-graining.
This matter is out of the scope of the current paper.
We will investigate here, the more general question: given a set of coarse-grained variables, $X$, how to describe its dynamics?
The question whether one always can find a suitable set of coarse-grained variables such that the Poisson-structure is found will be addressed in a future paper \cite{Pet08_2}.

The method used is the projection operator approach to non-equilibrium statistical mechanics.
This approach is related to topics such as BBGKY-hierarchy and Green-Kubo relations.
The projection operator method followed from a program of formal solutions of the Liouville equation starting from the middle of the 20th century.
Different mathematical techniques have been used, such as diagrammatic \cite{Pri61}, continued fractions \cite{Mor65_2}, Hilbert-space approaches \cite{Mor65,Zwa01}.

One of the problems when trying to apply projection operator formalism is which one to use.
Since there are many flavors, starting with a different flavor will result in a different outcome.
The outcome of all the projection-operator theories is a formally exact result.
This means the Liouville equation cast in a different form (and slightly different among different flavors).
It might therefore seem that all are equally good starting points, since nothing is lost.
This is not the case.
The equations resulting from projection operator formalism are an alternative statement of a problem that can not be solved exactly.
Nothing is lost, but neither anything is gained.
So it seems.
The strength of the alternative statements is the possibility to make approximations.
The different flavors of projection operator formalism should therefore be judged by the quality of the approximations one can make.
The best flavor is the one that, after approximations, produces results that are closest to exact results.

We find that, what we call, the ``Zwanzig flavor'' is in this respect superior to the others.
We proof that many of the projection-operator methods produce results that are, after approximation, valid only near equilibrium and at the thermodynamics limit.
They are therefore no good basis for a general theory of non-equilibrium thermodynamics.
The main argument of why the Zwanzig formalism is superior comes from optimal-prediction theory \cite{Cho98,Cho00}.

One of the main contribution of this paper is in pointing out that the correct starting point for non-equilibrium thermodynamics is the ``Zwanzig flavor'' projection operator formalism.
From this it follows that {\em the} ensemble is the microcanonical ensemble.
The entropy definition follows from this well motivated choice.
Equilibrium statistical mechanics then follows from the definition of the entropy.
So, here, differently from other derivations, statistical mechanics follows from the derivation and is not used as an input.

The goal of the projection-operator formalism is to describe a system using less degrees of freedom.
This is coarse-graining.
Coarse-graining is the second main topic to be discussed.
We will provide a method for progressive coarse-graining (of already coarse-grained descriptions).
This includes issues such as memory effects.
We will derive useful expressions relating the quantities at different levels of coarse-graining.
Our approach to coarse-graining circumvents the requirement of a wide separation of time-scales.
We will gives expressions that are generally valid.

Our goal is to explicitly establish a useful theory that is valid outside of the thermodynamic limit and far from equilibrium.
However, to show that the results have the correct limiting behavior we will also discuss the near-equilibrium linear regime and the non-linear continuum limit.
In the linear regime these relations corresponds to those found in homogenization theory \cite{Cio99}.

A specific form of coarse-graining occurs when locally the thermodynamic-limit is valid.
In this case, generalized canonical ensembles, do appear.
A discussion on this topic will put our approach in a wider perspective.
We will briefly discuss how to relate this work to issues as equivalences of ensembles (large-deviation theory \cite{Ell99,Ell00}).
Also the relation with approaches starting from a Gibbs-entropy definition will be discussed.

\subsection{notation}

In this paper we will mainly use an index-free notation.
Most quantities can be interpreted as columns of numbers.
So for $A$, $B$, $X$ or $Y$ one could imagine $A^i$, $B^i$, $X^i$ or $Y^i$ with $i$ running from $1$ to $n$.
There is not a lot of ``structure'', such as a metric, defined.
The only relevant structure turns out to be the (pushforward of the) Liouville measure or volume form.
Therefore we will not assume that quantities are tensorial pers\'{e}.
Where necessary transformation rules will be provided.
Especially entropy will turn out not to be a scalar quantity.

In many equations we will use a dot-product.
The dot-product indicates a contraction over indexes.
However, since there is no metric defined, most contractions would not make sense.
The dot-product indicates allowed contractions.
It indicates a dual-pairing rather than an inner-product.
We will, often implicitly, assume a upper and lower index convention to distinguish between the dual spaces.
Contractions are only possible over upper and lower indexes of components defined on spaces which are each others duals.
To demonstrate this convention we give the placement of the indexes of some of the quantities to be encountered:
\begin{equation}
  X^i, \, A^i, \, M^{ij}, \,\Omega^{ij}, \,\Omega^{ij}, \, \Lambda_{ij}, (\sqrt{2M})^i_\alpha, (dW)^\alpha.
\end{equation}

This has as consequence that we write, e.g, for a coordinate transformation or coarse-graining, of $\Omega^{ij}$,
\begin{equation}
  \ten{\Omega}^y = \frac{\partial Y}{\partial X} \cdot \ten{\Omega} \cdot \frac{\partial Y}{\partial X}, \; \Omega^{y,i'j'} = \frac{\partial Y^{i'}}{\partial X^{i}} \cdot \Omega^{ij} \cdot \frac{\partial Y^{j'}}{\partial X^{j}}.
\end{equation}
Note that this convention is different from matrix multiplication because there one would expect an transposed matrix at the end.
The transposed operator in, e.g, $\Omega^T=-\Omega$ indicates interchanging of indexes, i.e., $\Omega^{ij}=-\Omega^{ji}$.
Note that interchanging of upper and lower indexes does not make much sense when no metric is defined.
So if one has a quantity $A^{ij}_k$ the transposition would automatically imply the interchange of indexes $i$ and $j$ since this is the only one allowed.
Taking into account these conventions the presented formulas are without ambiguities.

In this paper we will solely use partial derivatives, and no functional, ones.
The reason is that we believe that coarse-grained systems should be viewed as finite, by definition.
The goal of coarse-graining is to reduce the number of freedom form say $10^{23}$ to $10^6$.
Continuum theories are smoothed theories that are not valid below a certain length-scale.
We think, for the purpose to solve these equations using a numerical method is better to write down finite equations to start with, see \S\ref{sect:concl}.

\section{A review of projection operator methods}

There are different flavors of projection-operator formalism.
In this section we will provide a brief outline of the main flavors.
Detailed technical arguments why the Zwanzig method is to be preferred compared to the alternative methods are mainly found in \S\ref{sect:proj-op}.
A mathematical exposition of the alternative methods is provided for in appendix \ref{app:proj-op}.

The Hilbert-space approach is most popular nowadays and is used in, e.g., in mode-coupling theory to describe the glass-transition \cite{Rei05}.
The Hilbert-space referred to is not the space formed by state-vectors in quantum mechanics.
The vectors in the ``Hilbert-space'' are functions defined on the microscopic phase space.
A ``vector'', say $A(\Gamma)$, is a quantity on the microscopic phase space in the sense a numerical value is assigned to any micro state $\Gamma$.
The definition of a Hilbert-space requires an inner-product.
Within the framework the inner-product of two function $A(\Gamma)$ and $B(\Gamma)$ is the expectation value of the product of $A^* B$ with respect to a (generalized) canonical ensemble.

The Liouville operator acts on the vectors/functions in the Hilbert space.
It is often referred to as a super-operator \cite{Zwa01} since, in the quantum-mechanical setting, it acts on observables and not on state-vectors.
Once a Hilbert-space is formed the formalism proceeds in a similar way for the classical and quantum-mechanical case.

A collections of macro states defines a linear subspace of the Hilbert-space.
The base vectors are given by functions $X^i(\Gamma)$.
Here $X^i(\Gamma)$ is a macroscopic state corresponding to a microscopic state $\Gamma$.
A finite number of macroscopic states $i=1, \dots, n$ are considered.
The ``projection'' of a general vector now consists of an projection onto this subspace.

Although the Hilbert-space approach is attractive from a formal point of view it has drawbacks.
The most important drawback is that one needs to define an inner-product on the space as a starting point.
Irrespective of the definition of the inner-product chosen one will get a formally exact result.
Clearly a formal result is not necessarily a practically useful result.
One needs to address the matter of which choice gives practically useful results.
The usual starting point is to use a canonical ensemble for this.
The Hilbert-space formulation using the equilibrium canonical ensemble results in the Mori-formalism.

The conceptual difficulty with defining an inner-product on the Hilbert-space is partly circumvented by taking a different point of view.
A more physical point of view is to interpret the projections as expectation values.
This is the point of view taken by Grabert \cite{Gra82}.
As is shown in the appendix \ref{app:proj-op} choosing equilibrium canonical expectation values results in the Mori-formalism.

Clearly, when interested in non-equilibrium phenomena it does not make much sense to consider equilibrium expectation values.
If one tries to do describe projections as expectation values using generalized canonical ensembles, however, one runs into trouble.
To form valid projection operators linearizations have to be performed that seem unsatisfactory from a physical point of view.
This is pointed out in detail in the appendix \ref{app:proj-op}.
Proceeding non-the-less with the linearized canonical distribution a generalized Langevin type equation is obtained.

This ``Langevin'' equation contains a fluctuating term.
The step from a formally exact result, namely, to a practically useful result is made by replacing the fluctuating (deterministic) term by a stochastic process.
We will point out in the main text that the step of modeling this fluctuating term as a stochastic process is not allowed for the generalized canonical ensemble.
The derived stochastic equation is valid only near equilibrium.
A second critique is that in Grabert's approach the use of the generalized canonical ensemble is motivated from outside the theory.
It comes from statistical mechanics reasoning using a Gibbs entropy.
It is not clear, a priori, why this is the correct ensemble to use at a small scale where fluctuations are important.

The use of the generalized canonical equation is in a sense illogical when one also can define a micro-canonical one.
This choice was made in the historic derivation by Zwanzig of a generalized Fokker-Planck equation\cite{Zwa61}.
By making this choice many things fall into place.
The awkwardness in the derivation disappears.
The projections can be interpreted as conditional expectation values without applying linearizations.
The expectation values are in accordance with the optimal predication framework \cite{Cho98, Cho00}.
The fluctuating term has the correct properties that allow it to be approximated by a stochastic process.

Below we will turn the story around.
When performing a projection there are important reasons to use the micro-canonical ensemble.
The main reason is that it is the optimal choice for computing conditional expectation values.
To define a conditional expectation value one needs to have an invariant measure.
Classical mechanics provides this measure, namely, the Liouville measure.
Continuing from this stage one finds a fundamental definition of the entropy.
It turns out that $\exp[S(X)]$ is the Liouville measure of phase space per unit volume coarse-grained space, i.e., a density of states.
If one likes mathematical terminology, it can be defined more rigorously as a Radon-Nikodym derivative towards the Lebesgue measure in the coarse-grained space of the pushforward of the Liouville measure to that coarse-grained space.
Because of this definition entropy is not a scalar quantity ($\exp[S(X)]$ is a density).
Careful application of projection-operator formalism gives us equilibrium and non-equilibrium thermodynamics.

\section{The nonlinear Langevin equation}
\label{section:Langevin}

The derivation of the nonlinear Langevin equation using projection operator formalism can be found in many standard texts and papers \cite{Zwa01, Gra82}.
Here we provide a straightforward derivation.
It is inspired on a derivation given in \cite{Kaw73}.

\subsection{The Liouville operator}

For any physical quantity $A$ the time development is described by means of a Liouville operator $\mathcal{L}$, formally,
\begin{equation}
  \frac{d}{dt} A_t = \mathcal{L} A_t.
  \label{eq:def_ode}
\end{equation}
It has the formal solution,
\begin{equation}
  A_t = \exp[\mathcal{L} t] \, A_0.
  \label{eq:time_evolution}
\end{equation}
In classical mechanics the quantity $A_t$ is fully specified by the microscopic state, $\Gamma_t$, of the system, so $A_t(\Gamma_0)=A(\Gamma_t)$.

Also in quantum-mechanics observables are completely determined by the time evolution of the initial microscopic state of the system.
In that case $A_t$ would be an operator evolving according to the Heisenberg description of the time-evolution.
In both cases, one finds for a product of quantities, that
\begin{equation}
  \begin{split}
    (A \, B)_t &= A_t \, B_t \rightarrow \\
    \exp[\mathcal{L} t] \, (A_0 \, B_0) &= (\exp[\mathcal{L} t] A_0) \, (\exp[\mathcal{L} t] B_0) \rightarrow \\
    \mathcal{L} \, (A_t \, B_t) &= (\mathcal{L} \, A_t) \, B_t + A_t \, (\mathcal{L} \, B_t).
  \end{split}
  \label{eq:product_rule}
\end{equation}
Here the last product (or Leibniz) rule is derived by time-derivative of the preceding identity.
We will encounter the Leibniz or product rule several times.

The property that makes it very  useful is that,
\begin{equation}
  \exp[\mathcal{L} t] \, f(A_0) = f(\exp[\mathcal{L} t] A_0). 
  \label{eq:paths}
\end{equation}
is valid if the product-rule is valid (for holomorphic functions $f$).
This rule might seem evident, in classical mechanics, if one considers trajectories through phase-space.

One can, however, write many systems, e.g., partial differential equations (first order in time), in the formal way of eq.~\eqref{eq:time_evolution} introducing a more general Liouville operator.
In this case the product rule does not necessarily hold.
Also, the Liouville operators produced by projection-operator formalism do not automatically obey the product rule.

Often one is interested in an ensemble of microscopic systems or in, e.g., time averages of a quantity.
For these cases it is convenient to introduce a dual object, $\mu$, that weighs the microstates.
The pairing of a quantity $A$ and the dual $\mu$ will be denoted as $\langle A, \mu \rangle$, which gives a (column of) number(s).
Let an operator, say $\mathcal{L}$ (but it can be any operator), work on $A$ then the conjugated operator is defined as
\begin{equation}
  \langle \mathcal{L} A, \mu \rangle = \langle A, \mathcal{L}^\dag \mu \rangle
  \label{eq:conjugated_definition},
\end{equation}
As a consequence, by using a series expansion for the exponential,
\begin{equation}
  \langle \exp[\mathcal{L} t] A, \mu \rangle = \langle A, \exp[\mathcal{L}^\dag t] \mu \rangle.
  \label{eq:conjugated_exponentials}
\end{equation}
So, if $\langle A_t, \mu \rangle$ is interpreted as an expectation value with respect to an initial ensemble of microstates then
\begin{equation}
  \langle A_t, \mu \rangle = \langle A, \mu_t \rangle \text{ with } \mu_t = \exp[\mathcal{L}^\dag t] \mu.
  \label{eq:expectation_value}
\end{equation}
Here fixed microstates are weighted by an evolving ensemble.
This is similar to the Heisenberg and Schr\"{o}dinger picture in quantum mechanics.
We will mainly work in the ``Heisenberg'' picture evolving $A$ and weighting with respect to the initial states.

When operator, $\mathcal{L}$ is a derivation (i.e. obeys the product rule), then combining eq.~\eqref{eq:conjugated_definition} and the product rule, eq.~\eqref{eq:product_rule}, gives
\begin{equation}
  \begin{split}
    \langle (\mathcal{L} A) \, B, \mu \rangle &= \langle \mathcal{L} \, (A \, B) , \mu \rangle - \langle A \, \mathcal{L} B, \mu \rangle\\
    &=  \langle A \, B , \mathcal{L}^\dag \mu \rangle - \langle A \, \mathcal{L} B, \mu \rangle.
  \end{split}
\end{equation}
This will be used in the derivations of the decomposed dynamics below.

We will focus on the classical description.
In a classical mechanics setting the microscopic evolution can always be thought of as a trajectory through phase space, parametrized by $\Gamma_t$.
Up to now expression eq.~\eqref{eq:time_evolution} was a formal solution of eq.~\eqref{eq:def_ode}.
For points in phase-space the operator is well defined, since
\begin{equation}
  \Gamma_t = \exp[\mathcal{L} t] \, \Gamma_0.
\end{equation}
Quantities are functions from phase space, $\mathcal{S}$,  to $\Reals^n$.
For the quantity $A(\Gamma)$ we have the formal relation
\begin{equation}
  A_t(\Gamma_0) = \exp[\mathcal{L} t] A(\Gamma_0) = A(\Gamma_t).
  \label{eq:A_t-def}
\end{equation}
An ensemble of initial states can be characterized by a measure $\mu$. The expectation value of a quantity with respect to this measure is
\begin{equation}
  \langle A_t, \mu \rangle = \int A_t(\Gamma_0) \, d\mu[\Gamma_0].
  \label{eq:pairing}
\end{equation}
Since $\mathcal{L}$ does itself not depend on time, a shift of the time index with a value $-t$ and using definition eq.~\eqref{eq:A_t-def} gives
\begin{equation}
    \int A_t(\Gamma_0) \, d\mu[\Gamma_0] = \int A(\Gamma_t) \, d\mu[\Gamma_0] = \int A(\Gamma_0) \, d\mu[\Gamma_{-t}].
\end{equation}
By comparing this expression with the definition, eq.~\eqref{eq:conjugated_exponentials}, for a subset of phase space points $\mathcal{B}$ (taken from the $\sigma$-algebra corresponding to the measure) one obtains
\begin{equation}
  \exp[\mathcal{L}^\dag t] \, \mu[\mathcal{B}] = \mu[\mathcal{B}_{-t}].
  \label{eq:L-dag}
\end{equation}

Using this equation, and the chain-rule of differentiation, one finds that
\begin{equation}
  (\mathcal{L}A_t)(\Gamma_0) = \dot{A}_t(\Gamma_0) = \dot{\Gamma}_t \cdot \frac{\partial A(\Gamma_t)}{\partial \Gamma_t} = \dot{\Gamma}_0 \cdot \frac{\partial A_t(\Gamma_0)}{\partial \Gamma_0},
  \label{eq:chain-rule}
\end{equation}
if $A(\Gamma)$ is differentiable.
Performing partial integration of $\langle \mathcal{L}A, \mu \rangle$, or using eq.~\eqref{eq:L-dag} one finds
\begin{equation}
  \mathcal{L}^\dag \mu[\mathcal{B}] = \int_\mathcal{B} - \frac{\partial}{\partial \Gamma} \cdot \Bigl ( \dot{\Gamma} \, d\mu[\Gamma] \Bigr ).
\end{equation}

For classical mechanics, when the coordinates parametrizing phase space, $\Gamma^c$, are canonical, Liouville's theorem holds.
Liouville's theorem states that microscopic phase space is incompressible,
\begin{equation}
  \frac{\partial}{\partial \Gamma^c} \cdot \dot{\Gamma}^c = 0.
\end{equation}
Using this observation one can define the Liouville measure for the parametrization with variables $\Gamma$ by making a coordinate transformation of the canonical variables $\Gamma^c$ to $\Gamma$,
\begin{equation}
  d\mu_L[\Gamma] = d\Gamma_c = \det \biggl ( \frac{\partial \Gamma^c}{\partial \Gamma} \biggr ) \, d\Gamma,
\end{equation}
here $d\Gamma$ is used to denote the Lebesgue measure.
This measure gives the usual volume of a (hyper)cube.
For this Liouville measure we have that
\begin{equation}
  \mathcal{L}^\dag \mu_L = 0.
  \label{eq:null_space}
\end{equation}
Measures that obey this property are called invariant.
Therefore the Liouville measure $\mu_L$ is an invariant measure.
When monitoring the weight of a set $\mathcal{B}_t$ evolving in time according to $\mathcal{L}$ then,  using eq.~\eqref{eq:L-dag}, this weight is constant if the measure is invariant.

When starting from a measure $\mu_0$ and computing an expectation value by using both time averaging one finds an time-averaged measure
\begin{equation}
  \bar{\mu} = \lim_{T \rightarrow \infty}\frac{1}{T} \int_0^T \mu_t \, dt = \lim_{T \rightarrow \infty} \frac{1}{T} \int_0^T \exp[\mathcal{L}^\dag t] \, \mu_0 \, dt,
\end{equation}
If this average measure exists then for $T\rightarrow \infty$
\begin{equation}
  \mathcal{L}^\dag \bar{\mu} = \lim_{T \rightarrow \infty} \frac{1}{T} \int_0^T \frac{d}{dt}\exp[\mathcal{L}^\dag t] \, \mu_0 \, dt = \lim_{T \rightarrow \infty}\frac{\mu_t-\mu_0}{T}=0,
\end{equation}
so $\bar{\mu}\in \mathrm{Null}(\mathcal{L}^\dag)$.
If eq.~\eqref{eq:null_space} has a unique solution (up to a multiplying constant), i.e., the null-space of $\mathcal{L}^\dag$ is one-dimensional, then this measure is necessary equal to $\bar{\mu}$.
This uniqueness of the time-average, irrespective of the initial condition is called ergodicity.
Because of the existence of conserved quantities such as energy, trajectories in phase space always remains inside a subspace.
Therefore ergodicity is usually interpreted as uniqueness of the solution of eq.~\eqref{eq:null_space} on the subspaces defined by conserved quantities.

Some kind of ergodicity is generally believed to be important for the fundamentals of statistical mechanics.
There are, however, many difficulties with this.
Moreover, here, we are interested in dynamic behavior.
Therefore very long time-intervals can not be considered.
One of the main goals of the present paper is to proceed as far as possible without making ergodicity assumptions.

\subsection{Decomposition of the Dynamics}

Let the ``real'' dynamics of the system be generated by a Liouville operator $\mathcal{L}$.
Let us now consider a quantity $A^\mathrm{fluct}_t$ that follows this dynamics approximately.
The difference between the ``real'' time evolution of an initial state given by $\exp[\mathcal{L} t] \, A^\mathrm{fluct}_0$, and $A^\mathrm{fluct}_t$ is,
\begin{multline}
    \exp[\mathcal{L} t] \, A^\mathrm{fluct}_0 - A^\mathrm{fluct}_t = \int_0^t \frac{d}{d{t'}} \Bigl( \exp[\mathcal{L} {t'}]\, A^\mathrm{fluct}_{t-{t'}}  \Bigr)\, d{t'}\\
    = \int_0^t \exp[\mathcal{L} {t'}]\, \Bigl(\mathcal{L}+ \frac{d}{d{t'}} \Bigr)  A^\mathrm{fluct}_{t-{t'}} \, d{t'}.
  \label{eq:decomposed_dynamics}
\end{multline}
We use the superscript ``fluct'' to indicate fluctuating or rapid dynamics, to be distinguished from the slow dynamics.

Within projection operator formalism the Liouville operator is decomposed as
\begin{equation}
  \mathcal{L} = \mathcal{P} \mathcal{L} + \mathcal{Q} \mathcal{L}.
\end{equation}
In the derivations occurring in the body of this paper we use the definitions,
\begin{equation}
  A^\mathrm{fluct}_t = \exp[\mathcal{Q} \mathcal{L} t] \, A_0 \text{ and }   A_t = \exp[\mathcal{L} t] \, A_0.
  \label{eq:fluct_tilde_def}
\end{equation}
Inserting this equation into eq.~\eqref{eq:decomposed_dynamics} one obtains
The decomposed equation using this quantity is
\begin{equation}
  \begin{split}
    \Delta A_t & = \int_0^t \exp[\mathcal{L} {t'}]\, \mathcal{P} \mathcal{L}  A^\mathrm{fluct}_{t-{t'}} \, d{t'}+ \Delta A^\mathrm{fluct}_t\\
    &= \int_0^t \exp[\mathcal{L} {t'}]\, \mathcal{P} \mathcal{L}  A_0 \, d{t'} \\ &\quad + \int_0^t \exp[\mathcal{L} {t'}]\, \mathcal{P} \mathcal{L}  \Delta A^\mathrm{fluct}_{t-{t'}} \, d{t'}+ \Delta A^\mathrm{fluct}_t
  \end{split}
  \label{eq:generalized_langevin_2}
\end{equation}
where $\Delta A^\mathrm{fluct}_t = A^\mathrm{fluct}_t- A^\mathrm{fluct}_0$ (similarly $\Delta A_t=A_t-A_0$).

The fluctuating dynamics is a solution of
\begin{equation}
  \frac{d}{dt} A^\mathrm{fluct}_t = \mathcal{Q} \mathcal{L} \, A^\mathrm{fluct}_t,
  \label{eq:fluctuating_dynamics}
\end{equation}
with initial value $A^\mathrm{fluct}_0 = A_0$.
An alternative definition, often found in literature, is to take an initial condition $\tilde{A}^\mathrm{fluct}_0 = \mathcal{Q} \, A_0$.
We will denote this definition using a tilde.
The solution is then, 
\begin{equation}
  \tilde{A}^\mathrm{fluct}_t = \exp[\mathcal{Q} \mathcal{L} t] \, \mathcal{Q} \, A_0.
  \label{eq:fluctuating_dynamics_solution}
\end{equation}
For this definition one finds the decomposition,
\begin{multline}
    A_t = \exp[\mathcal{L} t] \, \mathcal{P} \, A_0 \quad + \\ \int_0^t \exp[\mathcal{L} s]\, \mathcal{P} \mathcal{L}  \tilde{A}^\mathrm{fluct}_{t-s} \, ds+ \tilde{A}^\mathrm{fluct}_t.
    \label{eq:generalized_langevin}
\end{multline}

Within the projection operator formalism the operators $\mathcal{P}$ and $\mathcal{Q}$ are projection operators.
They have the properties,
\begin{equation}
  \mathcal{P} = \mathcal{P}^2 , \; \mathcal{Q} = \mathcal{Q}^2 \text{ and } \mathcal{P}+\mathcal{Q}=1.
\end{equation}
By convention $\mathcal{P}$ is supposed to filter-out the slow (or ``relevant'' dynamics) and $\mathcal{Q}$ the fast (fluctuating or ``irrelevant'') dynamics.

As a consequence of the dynamics and the initial value, combined with the projection-property of the operator one finds
\begin{equation}
  \mathcal{P} \, \tilde{A}^\mathrm{fluct}_t = 0.
  \label{eq:projected_fluctuations}
\end{equation}
For the fluctuating contributions as we have defined them the property is a bit weaker, namely,
\begin{equation}
  \mathcal{P} A^\mathrm{fluct}_t=A_0, \text{ such that } \mathcal{P} \Delta A^\mathrm{fluct}_t = 0.
  \label{eq:projected_fluctuations_2}
\end{equation}

The purpose of the projection operators is to filter out ``irrelevant'' contributions.
In coarse-grained descriptions one wants to describe the problem in terms of coarse-grained variables that have the property $X=\mathcal{P} X$.
This means they are invariant under projection.

In most derivations of projection-operator formalisms $\tilde{A}^\mathrm{fluct}_t$ instead of $A^\mathrm{fluct}_t$ is used.
Note that the equation for $A=X$ as given by eq.~\eqref{eq:generalized_langevin} is trivial.
Since $\tilde{X}^\mathrm{fluct}_0 = \mathcal{Q} \, X=0$ one finds that $\tilde{X}^\mathrm{fluct}_t=0$.
Therefore the usual approach to obtain an equation for the change of $X$ is to consider first $A=\dot{X}=\mathcal{L} \, X$.
Next, to obtain a change of $X$ one integrates $\dot{X}$ over a time-interval.
An alternative approach is to consider eq.~\eqref{eq:generalized_langevin_2} with $A=X$.
This will give exactly the same equation without the need for the time-integration.
Also eq.~\eqref{eq:generalized_langevin_2} is a little bit more convenient as starting point for an approximation using stochastic processes.

In the derivations presented in appendix \ref{app:proj-op} that compares we will consistently use $\tilde{A}^\mathrm{fluct}_t$ instead of $A^\mathrm{fluct}_t$.
The main reason is that the Robertson/Grabert approach is difficult to express otherwise.
In the body of the paper where we focus on the Zwanzig operator formalism using $A^\mathrm{fluct}_t$ turn out to be more convenient.

\subsection{Problems of the canonical-based formalisms}
\label{sect:proj-op}

To proceed further one needs to define a projection operator, $\mathcal{P}$.
In this paper we will argue that the projection operator as defined by Zwanzig is to be preferred.
Other versions can be seen as near equilibrium, or thermodynamic limit, approximations.
Derivations usually start from the (generalized) canonical ensemble.
This starting point is understandable from the point of view of statistical mechanics.
The canonical ensemble is much easier to handle.

The derivations based on this approximation we have labeled Mori, Robertson and Grabert derivations.
The derivations, are outlined in appendix \ref{app:proj-op}.
Before we start with discussing the Zwanzig formalism we want to discuss the properties of these other flavors that make them unsuited as a starting point for a general framework of coarse-graining.
The core problem with these derivations is that one micro state $\Gamma$ is associated to a macro state, $X$, via a many-to-one transformation $X(\Gamma)$.
The canonical ensemble, however, associates one micro state $\Gamma$ to more than one macro state $X$.
The generalized canonical ensemble used to define a conditional measure, $\mu^\mathrm{rel}(X)[\mathcal{B}]$ associates a significant weight to not only the microstates with $X(\Gamma)=X$, but for any $\Gamma$ with $X(\Gamma)$ near to $X$.
So a micro state contributes to more than one macro state $X$.

One can try to define a projection as a conditional expectation value, $\mathrm{P}^\mathrm{C} X \equiv \langle X, \mu^\mathrm{rel}(X)\rangle$.
By construction this operator has the property,
\begin{equation}
  \mathcal{P} \, X = X.
  \label{eq:linear-projection}
\end{equation}
However, for general nonlinear functions of $g(X)$,
\begin{equation}
  (\mathcal{P} \, g)(X) \ne g(X).
  \label{eq:nonlinear-projection}
\end{equation}
The reason is the mentioned asymmetry in association between microstates $X$ and macrostates $\Gamma$.

For the canonical ensemble, applying the conditional expectation value several times gives $\mathrm{P}^\mathrm{C}\mathrm{P}^\mathrm{C} \ne \mathrm{P}^\mathrm{C}$.
Therefore the canonical expectation value does not define a projection.
This is a consequence of eq.~\eqref{eq:nonlinear-projection}.
The projection property is important because it results in eq.~\eqref{eq:projected_fluctuations} and eq.~\eqref{eq:projected_fluctuations_2}.
If $\mathrm{P}^\mathrm{C}$ is used the ``projection'' of the fluctuating contribution is this non-zero.
Its expectation value (according to the canonical ensemble) is non-zero.

Within the Mori and the Robertson-Grabert formalism the ``projection'' property is restored by applying a linearization.
Here,
\begin{equation}
  (\mathcal{P} \, g)(X) = g(x^\mathrm{eq}) + (X-x^\mathrm{eq}) \cdot \frac{\partial g(x^\mathrm{eq})}{\partial x^\mathrm{eq}}.
  \label{eq:nonlinear-linearized}
\end{equation}
Because of property eq.~\eqref{eq:linear-projection} applying the projection many times gives the same result.
Therefore the Mori and Robertson/Grabert projections are genuine projection operators.
The applied linearization makes that, if one views the projection as taking an expectation value, the used ensemble is not canonical anymore.
The status is now that for the fluctuations eq.~\eqref{eq:projected_fluctuations} is obeyed.
So, the expectation value of the fluctuations is zero.
However, one can doubt much is gained by the fact the expectation value is zero with respect to a non-canonical ensemble.

Within the Mori approach the equilibrium value $x^\mathrm{eq}$ is fixed.
The Mori formalism is arrived at by means of a linearization.
As already mentioned it can be restated in a Hilbert-space formalism \cite{Zwa01}.
Expectation values using the (generalized) canonical ensemble is than used to define an inner product.
The inner product is used to define a projection.
The macroscopic variables $X^i(\Gamma)$ are vectors in Hilbert-space that define a subspace where is projected on.
This mathematical elegance is one of the reasons for the popularity of the Mori-formalism.

In the Robertson and Grabert approaches the value of $x^\mathrm{eq}$ is made to change in time.
The Robertson and Grabert approaches can be seen as damage-control because of the loss of the canonical ensemble.
They reduce the error made by the linearization.
This is at the expense of complicating the framework by the need to introduce a time-dependent projection operator, see appendix \ref{app:proj-op}.

The treatments of Mori, Robertson and Grabert restore the projection property.
Because of this the framework of the projection operator method can be used.
Therefore formally exact generalized Langevin equations can be derived.
However, the non-equality eq.~\eqref{eq:nonlinear-projection} remains there.
As we will argue later, eq.~\eqref{eq:nonlinear-projection}, excludes the formalisms as a basis for further (stochastic) approximations.
This fact seems to be missed in most of the literature.

The main shortcoming of the Mori formalism, eq.~\eqref{eq:nonlinear-projection}, can also be elevated in an alternative way (i.e., different from the Robertson and Grabert ``improvements'').
This is by extending the basis of the subspace.
One example, of such an extension, is not only to use functions, $X^i(\Gamma)$, but also quadratic ones, $X^i(\Gamma)\, X^j(\Gamma)$.
Clearly in this way an equality is obtained, instead of the inequality eq.~\eqref{eq:nonlinear-projection}, for not only linear functions but also quadratic ones.
To suite a particular problem one can use any basis.
This choice of basis makes the Mori formalism very versatile.

Many times the different projection operator formalisms are presented as a choice dictated by convenience.
In analyzing different methods we came to the conclusion that these are alternative methods to battle the consequences of eq.~\eqref{eq:nonlinear-projection} (while keeping the projection property).
The one thing in common is that the starting point is always the (generalized) canonical ensemble.
All have also in common that the projection operator obtained can not be interpreted as an exact expectation value using this ensemble.

\section{The Zwanzig formalism}

The Zwanzig formalism \cite{Zwa61}, which was historically the first projection operator formalism introduced, uses the generalized microcanonical ensemble instead of the (generalized) canonical one.
One could also view this as just a choice one can make.
From statistical mechanics one might have the idea that there is little difference because of the equivalence of ensembles.
This is all true, but only in the thermodynamic limit and close to equilibrium.

We will turn the story inside-out.
Because of the central importance of
\begin{equation}
  (\mathcal{P} \, g)(X) = g(X),
  \label{eq:optimal-projection}
\end{equation}
let us try to find a projection operator formalism that obeys this equality.
This turns out to be the Zwanzig operator formalism and it turns out to use the (generalized) microcanonical ensemble!

The big picture is as follows.
Starting from a theory of coarse-graining, i.e. projection from the micro state to a macro state one finds the microcanonical ensemble.
This theory gives a generalized Langevin equation that has superior qualities compared to the Mori one (and its extensions) because of eq.~\eqref{eq:optimal-projection}.
Coarse-graining therefore dictates the microcanonical ensemble and from this the microcanonical definition of entropy.
Therefore, also statistical mechanics follows from this theory of coarse-graining and not the other way around.
The transition from the microcanonical ensemble to the canonical follows from large-deviation theory \cite{Rue69, Lan73, Ell99}.

To find a projection operator that obeys eq.~\eqref{eq:optimal-projection} we will take the point of view of optimal prediction theory \cite{Cho98, Cho00}.
Here we will start out with a general measure $\mu$, i.e., not necessarily the Liouville measure.
Using this measure we will try to find an optimal prediction.
Say we want to find a prediction or projection of $A(\Gamma)$ by means of a function $f(X(\Gamma))$ that depends on $\Gamma$ through $X$.
The prediction is said to be optimal (with respect to $\mu$) when
\begin{equation}
  \langle |A-f(X)|^2, \mu \rangle = \text{minimal}.
  \label{eq:optimal_prediction}
\end{equation}
This optimum is given by $f(X)=\mathbb{E}_\mu(A|X)$, the conditional expectation value.
It has the defining property that for any function $g(X(\Gamma))$,
\begin{equation}
  \langle g(X) \, \bigl(A- \mathbb{E}_\mu(A|X)\bigr) , \mu \rangle=0.
  \label{eq:conditional_prob_property}
\end{equation}
Using this property the cross term cancels in:
\begin{multline}
    \langle |A-f(X)|^2, \mu \rangle= \langle |A-\mathbb{E}_\mu(A|X)|^2, \mu \rangle\\
    + 2 \mathrm{Re}   \langle (\mathbb{E}_\mu(A|X)-f(X))^* \, (A-\mathbb{E}_\mu(A|X) , \mu \rangle   \\
    + \langle |\mathbb{E}_\mu(A|X) -f(X) |^2, \mu \rangle\\
    = \langle |A-\mathbb{E}_\mu(A|X)|^2, \mu \rangle + \langle |\mathbb{E}_\mu(A|X) -f(X) |^2, \mu \rangle
\end{multline}
and that therefore $f(X)=\mathbb{E}_\mu(A|X)$ is indeed optimal.
In a similar way one can proof it is unique (in a square integrable sense).
Clearly, from eq.~\eqref{eq:optimal_prediction} follows that if $A$ is a function of $X$, the optimum is just $\mathbb{E}_\mu(A|X)=A(X)$.
Therefore for any $\mathbb{E}_\mu(\mathbb{E}_\mu(A|X)|X)=\mathbb{E}_\mu(A|X)$.
This is the defining property of a projection operator
\begin{equation}
  (\mathcal{P}_\mu A)(X) \equiv \mathbb{E}_\mu(A|X)
  \label{eq:Zwanzig_projection}
\end{equation}
where $\mu$ is a measure on the microscopic space.
This is the Zwanzig projection operator.
The Zwanzig projection operator obeys eq.~\eqref{eq:optimal-projection} for general functions $g(X)$, whereas Mori and Robertson/Grabert projection operators do not.
The optimality of the Zwanzig projection is a strong reason for preferring it.
The value does depend, however, on which measure $\mu$ is used.
The conclusion here is for a specified measure, say $\mu_L$, the Zwanzig projection give the optimal projection, better than, e.g., the Mori or Robertson/Grabert projection.

The definition of eq.~\eqref{eq:conditional_prob_property} is implicit.
It can be made more explicit by introducing an indicator function or a Dirac measure.
For a subset $\mathcal{B}$ of the coarse-grained $X$ space we define the indicator function and the Dirac measure as follows
\begin{equation}
  \mathbf{1}_\mathcal{B}(X) = \delta_X[\mathcal{B}] = 
\left\{\begin{matrix} 
1 &\mbox{if}\ X \in \mathcal{B}, \\
0 &\mbox{if}\ X \notin \mathcal{B}.
\end{matrix}\right.
\end{equation}
Having a measure on the microscopic space one can define a pushforward measure on the coarse-grained space as
\begin{equation}
  X_*(\mu)[\mathcal{B}] = \mu[X^{-1}(\mathcal{B})] = \int \mathbf{1}_\mathcal{B}(X(\Gamma)) \, d\mu[\Gamma].
\end{equation}
Taking $g(X)=\mathbf{1}_\mathcal{B}(X)$ in eq.~\eqref{eq:conditional_prob_property} one has,
\begin{multline}
  \int \mathbf{1}_\mathcal{B}(X(\Gamma)) \, A(\Gamma) \, d\mu[\Gamma] = \\ \int \mathbf{1}_\mathcal{B}(X(\Gamma)) \, \mathbb{E}_\mu(A|X(\Gamma)) \, d\mu[\Gamma] =\\ \int_\mathcal{B} \mathbb{E}_\mu(A|X) \, dX_*(\mu)[X].
\end{multline}
Replacing the indicator function by the Dirac measure one can rewrite this as
\begin{equation}
  \int_\mathcal{B} \mathbb{E}_\mu(A|X) \, dX_*(\mu)[X] = \int \delta_{X(\Gamma)}[\mathcal{B}] \, A(\Gamma) \, d\mu[\Gamma].
  \label{eq:expectation-value_def}
\end{equation}

If the relation $X(\Gamma)$ and the measure $\mu$ are smooth enough (more precisely $X_*(\mu)$ is absolutely continuous with respect to the Lebesgue measure on the $X$-space) we can introduce a, so called, Radon-Nikodym derivative that defines the entropy as
\begin{equation}
  \exp[S_\mu(X)] = \frac{dX_*(\mu)[X]}{dX}.
  \label{eq:entropy_definition}
\end{equation}
Here we use the convention $dX=d\mu_\mathrm{Lebesgue}[X]$.
Taking the derivative left and right-hand-side of eq.~\eqref{eq:expectation-value_def} gives
\begin{multline}
  \mathbb{E}_\mu(A|X) \, \frac{dX_*(\mu)[X]}{dX} = \int \frac{d\delta_{X(\Gamma)}[X]}{dX} \, A(\Gamma) \, d\mu[\Gamma]\\
  \mathbb{E}_\mu(A|X) = \exp[-S_\mu(X)] \, \int \delta[X(\Gamma)-X] \, A(\Gamma) \, d\mu[\Gamma],
\end{multline}
where,
\begin{equation}
  \delta[\tilde{X}-X] \equiv \frac{d\delta_{\tilde{X}}[X]}{dX},
  \label{eq:delta-function}
\end{equation}
is the Dirac delta-function that should be interpreted as a distribution rather than a function on the $X$-space.

The projection operator, eq.~\eqref{eq:Zwanzig_projection}, can be seen as averaging the function $A(\Gamma)$ over the subspace defined by $X(\Gamma)=X$.
In the derivation (by means of introducing a Radon-Nikodym derivative) we assumed that $X(\Gamma)$ is a continuous function.
In many cases macroscopic variables are discrete, e.g., the number of particles present in a cell in space.
We will ignore this case in this paper and assume that all quantities are continuous.

The quantity $\exp[S_\mu(X)]$ is the microscopic phase space measure, $\mu$, per unit Lebesgue measure of macroscopic space.
The convenience of introducing the Lebesgue measure is that it is translational invariant.
Using this pairing notation of quantities and their duals we can rewrite eq.~\eqref{eq:Zwanzig_projection} as,
\begin{equation}
  \begin{split}
    \mathbb{E}_\mu(A|X) &= \exp[-S_\mu(X)] \langle A \, \delta[X-X(\Gamma)] , \mu \rangle\\
    &= \langle A , \mu^\mathrm{rel,Z}(X) \rangle,
  \end{split}
  \label{eq:conditional_expectation}
\end{equation}
where, at least formally,
\begin{equation}
  d\mu^\mathrm{rel,Z}(X)[\Gamma] =  \exp[-S_\mu(X)] \, \delta[X-X(\Gamma)] \, d\mu[\Gamma],
  \label{eq:microcanonical}
\end{equation}
defines a microcanonical ensemble corresponding to macro state $X$.

Let us now develop a generalized Langevin equation for the Zwanzig projection operator.
So we will develop the terms arising in eq.~\eqref{eq:generalized_langevin_2} for the case of the optimal-prediction/Zwanzig projection.
The first term
\begin{equation}
  \exp[\mathcal{L}t] \, \mathcal{P}_\mu \, A_0 = \exp[\mathcal{L}t] \, \mathbb{E}_\mu(A|X_0) = \mathbb{E}_\mu(A|X_t),
\end{equation}
can be interpreted as the conditional expectation of $A$ with respect to the current macro state $X_t$.

The integrand in the second term in the last equation of eq.~\eqref{eq:generalized_langevin_2} becomes,
\begin{equation}
  \exp[\mathcal{L} {t'}]\, \mathcal{P}_\mu \mathcal{L}  \Delta A^\mathrm{fluct}_{t-{t'}} = \mathbb{E}_\mu( \mathcal{L}  \, \Delta A^\mathrm{fluct}_{t-{t'}} |X_{t'}).
\end{equation}
The combination $\mathcal{L}  \, A^\mathrm{fluct}_{t-{t'}}$ is a bit problematic, because the generator of the dynamics for $A^\mathrm{fluct}$ is $\mathcal{Q}\mathcal{L}$ not $\mathcal{L}$ (see eq.~\eqref{eq:fluctuating_dynamics}).

Using eq.~\eqref{eq:Zwanzig_projection} with the delta-function definition, eq.~\eqref{eq:delta-function}, inserted gives for the enumerator,
\begin{multline}
  \langle (\mathcal{L}  \, \Delta A ^\mathrm{fluct}_{t-{t'}}) \, \delta[X-X(\Gamma)] , \mu \rangle = \\
  - \langle  \Delta A ^\mathrm{fluct}_{t-{t'}} \, \mathcal{L} \, \delta[X-X(\Gamma)] , \mu \rangle\\ 
  + \langle  \Delta A^\mathrm{fluct}_{t-{t'}} \, \delta[X-X(\Gamma)] , \mathcal{L}^\dag \mu \rangle.
  \label{eq:zwanzig_time_correlation}
\end{multline}
Simplifying the first term on the right-hand-side gives,
\begin{multline}
  - \langle  \Delta A^\mathrm{fluct}_{t-{t'}} \, \mathcal{L} \, \delta[X-X(\Gamma)] , \mu \rangle = \\
  \frac{\partial}{\partial X} \cdot \langle  \dot{X}_0 \, \Delta{A}^\mathrm{fluct}_{t-{t'}} \, \delta[X-X(\Gamma)] , \mu \rangle.
\end{multline}
Putting also the denominator in eq.~\eqref{eq:Zwanzig_projection} back into place gives,
\begin{multline}
  \mathbb{E}_\mu( \mathcal{L}  \, \Delta A^\mathrm{fluct}_{t-{t'}} |X_{t'}) = \exp[-S_\mu(X_{t'})] \times \\
  \frac{\partial}{\partial X_{t'}} \cdot \Bigl ( \exp[S(X_{t'})] \, \mathbb{E}_\mu( \dot{X}_0 \, \Delta {A}^\mathrm{fluct}_{t-{t'}} |X_{t'})\Bigr)\\ + \langle  \Delta {A}^\mathrm{fluct}_{t-{t'}} \, \delta[X-X(\Gamma)] , \mathcal{L}^\dag \mu \rangle.
  \label{eq:zwanzig_time_correlation_2}
\end{multline}

Here the last term is very inconvenient.
It disappears if one uses an invariant measure, such that $\mathcal{L}^\dag \mu=0$.
Note that it is not required that the measure is ergodic here.
Since the Liouville measure is known, a priori, to be invariant it is a logical and convenient choice to take: $\mu=\mu_L$.
For this choice we will define
\begin{equation}
    \mathbb{E}(A|X) \equiv \mathbb{E}_{\mu_L}(A|X), \; 
    S(X) \equiv S_{\mu_L}(X).
\end{equation}
The relation \eqref{eq:zwanzig_time_correlation_2} can also be used to derive a, so-called, degeneracy relation.
For $\Delta {A}^\mathrm{fluct}=1$ one obtains 
\begin{equation}
     \frac{\partial}{\partial X} \cdot \Bigl ( \exp[S(X)] \, \mathbb{E}(\dot{X} |X) \Bigr) = 0.
    \label{eq:degeneracy}
\end{equation}
This can be read as a generalization of the Liouville theorem to the coarse-grained case.

By taking $\dot{X}^\mathrm{fluct} = \mathcal{Q} \, \dot{X} = \dot{X} - \mathbb{E}(\dot{X}|X)$ and noting that $\mathbb{E}(\mathbb{E}(A|X)\,B|X_s)=\mathbb{E}(A|X_s) \, \mathbb{E}(B|X_s)$ one finds that
\begin{equation}
  \mathbb{E}( \dot{X}_0 \, \Delta A^\mathrm{fluct}_{t-{t'}} |X_{t'}) = \mathbb{E}( \dot{X}_0^\mathrm{fluct} \, \Delta A^\mathrm{fluct}_{t-{t'}} |X_{t'}),
\end{equation}
because $\mathbb{E}(\Delta A^\mathrm{fluct}_{t-{t'}} |X)=0$ according to eq.~\eqref{eq:projected_fluctuations_2}.
Summarizing:
\begin{multline}
  \mathbb{E}( \mathcal{L}  \, \Delta A^\mathrm{fluct}_{t-{t'}} |X_{t'}) = \exp[-S(X_{t'})] \times\\
 \frac{\partial}{\partial X_{t'}} \cdot \Bigl ( \exp[S(X_{t'})] \, \mathbb{E}(\dot{X}^\mathrm{fluct}_0 \, \Delta A^\mathrm{fluct}_{t-{t'}} |X_{t'}).
 \label{eq:dissipative_term}
\end{multline}

The full non-linear Langevin equation for the Zwanzig-projection is
\begin{widetext}
  \begin{equation}
    \Delta A_t = \int_0^t \mathbb{E}(\dot{A}|X_{t'}) \, d{t'} + \int_0^t \exp[-S(X_{t'})] \, \frac{\partial}{\partial X_{t'}} \cdot \Bigl ( \exp[S(X_{t'})] \, \mathbb{E}(\dot{X}^\mathrm{fluct}_0 \, \Delta A^\mathrm{fluct}_{t-{t'}} |X_{t'}) \Bigr ) \, d{t'} 
    + \Delta A^\mathrm{fluct}_t.
    \label{eq:nonlinear_langevin_2}
  \end{equation}
  Using $A=X$ one obtains an equation that can be used as an starting point to obtaining approximate coarse-grained equations for $X$.
  \begin{equation}
    \Delta X_t = \int_0^t \mathbb{E}(\dot{X}|X_{t'}) \, d{t'}+ \int_0^t \exp[-S(X_{t'})] \, \frac{\partial}{\partial X_{t'}} \cdot \Bigl ( \exp[S(X_{t'})] \, \tilde{\ten{M}}_{t-t'}^T(X_{t'}) \Bigr ) \, d{t'} + \Delta X^\mathrm{fluct}_t,
    \label{eq:langevin_X}
  \end{equation}
\end{widetext}
where
\begin{equation}
  \begin{split}
    X^\mathrm{fluct}_t &= \exp[\mathcal{Q} \mathcal{L} \, t] \, X_0\\
    \tilde{\ten{M}}_{\tau}(X) &= \mathbb{E}(\Delta X^\mathrm{fluct}_{\tau} \dot{X}^\mathrm{fluct}_0|X).
  \end{split}
\end{equation}

The importance of the Zwanzig projection formalism for coarse-graining is that the fluctuating dynamics is split of.
The time auto-correlations can be obtained by purely considering the fluctuating dynamics.
One can solve the fluctuating dynamics and use the result as input to model the dynamics of $X$.
The art of coarse-graining is to make a good choice for $X$.
The $X$ should be, preferably, chosen such that the fluctuating dynamics is fast.
This means that $X^\mathrm{fluct}_t$'s decorrelate quickly.
When this is the case the fluctuations can be well approximated by means of a stochastic variable.

An important property of the Zwanzig projection is the following. 
For quantities, $A$, that can be expressed as a function of $X$ (so $A(\Gamma)=A(X(\Gamma))$) one has
\begin{equation}
  \mathcal{P} (A \, B) (X)= \mathbb{E}(A\, B|X) = A(X) \, \mathbb{E}(B|X).
  \label{eq:cond-expect-prop}
\end{equation}
This is even a bit stronger than property eq.~\eqref{eq:optimal-projection}.
Now consider two quantities $A(X)$ and $B(X)$ that depend on the macro state $X$.
For these we have
\begin{equation}
  \begin{split}
    \mathcal{P} \mathcal{L} (A \, B) (X) &= \mathbb{E}(\mathcal{L} (A\, B)|X) \\
    &= \mathbb{E}((\mathcal{L} A) \,  B + A \, (\mathcal{L} B)|X)\\
    &= \mathbb{E}(\mathcal{L} A |X) B(X) + A(X) \mathbb{E}(\mathcal{L} B |X)\\
    \mathcal{P} \mathcal{L} (A \, B) &= (\mathcal{P} \mathcal{L} A) B + A (\mathcal{P} \mathcal{L} B)
  \end{split}
\end{equation}
This result is obtained by combining the product rule for $\mathcal{L}$, eq.~\eqref{eq:product_rule} and eq.~\eqref{eq:cond-expect-prop}.
This last line tells is that the combined operator $\mathcal{P} \mathcal{L}$ obeys the product rule when acting on functions of $X$.
By subtracting the equality from eq~\eqref{eq:product_rule} one also has
\begin{equation}
  \mathcal{Q} \mathcal{L} (A \, B) = (\mathcal{Q} \mathcal{L} A) B + A (\mathcal{Q} \mathcal{L} B).
\end{equation}
So, $\mathcal{Q} \mathcal{L}$ obeys the product rule and can therefore be interpreted as a derivation!
This gives, according to eq.~\eqref{eq:paths}, that for any (holomorphic) function $A(X)$,
\begin{equation}
  A^\mathrm{fluct}_t = A(X^\mathrm{fluct}_t).
  \label{eq:coarse_grained_product_rule}
\end{equation}
Therefore, in the Zwanzig formalism, the fluctuating dynamics can be imagined as a trajectory through macroscopic phase space.
This is certainly not the case for any of the other projection operator formalisms (Mori, Robertson, Grabert).
Because for these formalism eq.~\eqref{eq:optimal-projection} is not found to be valid.
This gives that in those cases, even when $A(X)$, that $A^\mathrm{fluct}_t \ne A(X^\mathrm{fluct}_t)$.

\section{Stochastic differential equations}

The generalized Langevin equation, Eq.~\eqref{eq:langevin_X}, is a formal decomposition of the microscopic equations of motion.
It contains no new information.
Full expressions of the fluctuating term $X^\mathrm{fluct}$ are very complicated.
Its use lies in the fact that it can be used as a starting point for approximations.

Suitable choices for the macroscopic variables $X$ can be made.
The usual approach is to choose the variables such the remainder characterized by $X^\mathrm{fluct}$ decorrelates quickly.
So, on the time that $X^\mathrm{fluct}_t$ decorrelates $X_t$ has barely moved.
Integrating Eq.~\eqref{eq:langevin_X} for $\Delta t$ (larger than the decorrelation time) and replacing functions dependencies of $X_{t'}$ by $X_0$ gives
\begin{multline}
  \Delta X  \approx \mathbb{E}(\dot{X}|X_0) \, \Delta t + \\
  \exp[-S(X_0)] \, \frac{\partial}{\partial X_0} \cdot \Bigl( \exp[S(X_0)] \,  \tilde{\ten{M}}^T(X_0)\Bigr ) \, \Delta t\\
  +  \Delta X^\mathrm{fluct}_t,
  \label{eq:Langevin_integrated}
\end{multline}
with
\begin{equation}
  \tilde{\ten{M}}(X_0) \approx \frac{1}{\Delta t} \int_{0}^{\Delta t} \mathbb{E}( \Delta X^\mathrm{fluct}_{t'} \dot{X}^\mathrm{fluct}_0 | X_0)\, d{t'} .
  \label{eq:dissipation}
\end{equation}
The modeling assumption is that (complete) decorrelation is very fast, i.e., the change of $X$ is very small on a time scale $\tau$.
One is interested in phenomena on time scales much larger than $\tau$.
Under the same assumptions one finds that for $\Delta t = \mathcal{O}(\tau)$,
\begin{multline}
  \mathbb{E}(\Delta X^\mathrm{fluct} \, \Delta X^\mathrm{fluct}|X_0) = \\ \int_{0}^{\Delta t} \Bigl( \mathbb{E}(\dot{X}^\mathrm{fluct}_{t'} (X^\mathrm{fluct}_t - X^\mathrm{fluct}_{t'}) |X_0) + \\ \mathbb{E}((X^\mathrm{fluct}_t - X^\mathrm{fluct}_{t'})\, \dot{X}^\mathrm{fluct}_{t'}  |X_0) \Bigr) \, d{t'} \\ \approx \int_{0}^{\Delta t} \Bigl( \mathbb{E}(\dot{X}^\mathrm{fluct}_0 (X^\mathrm{fluct}_{t-{t'}} - X^\mathrm{fluct}_0) |X_0) + \\ \mathbb{E}((X^\mathrm{fluct}_{t-{t'}} - X^\mathrm{fluct}_0)\, \dot{X}^\mathrm{fluct}_0  |X_0) \Bigr) \, d{t'} \\
  = (\tilde{\ten{M}}^T(X_0) + \tilde{\ten{M}}(X_0)) \, \Delta t \equiv 2 \, \ten{M}(X_0) \, \Delta t.
  \label{eq:symmetric_M}
\end{multline}
By definition, $\ten{M}$ is symmetric and positive (semi) definite.
However, $\tilde{\ten{M}}$ not necessarily so.
In the general case one can write
\begin{equation}
  \tilde{\ten{M}} = \ten{M}  + \mathbf{A},
\end{equation}
where $\mathbf{A}$ is anti-symmetric, i.e., $\mathbf{A}^T = -\mathbf{A}$.

Arguments like the central limit theorem (more specifically Donker's theorem) can be used to argue that $X^\mathrm{fluct}$ can be well approximated using a Wiener process $W$,
\begin{equation}
  \Delta X^\mathrm{fluct}_t \approx \sqrt{2 \ten{M}(X_0)} \cdot \Delta W_t,
  \label{eq:white_noise}
\end{equation}
where $\ten{M}$ is a positive definite matrix and $\Delta W_t = W_{t} - W_0$.
A Wiener process is a Gaussian stochastic process.
Each increment over a time-step $\Delta t$ has zero average and variance $\Delta t$,
\begin{equation}
  \mathbb{E}(\Delta W | X_0) = 0, \text{ and }  \mathbb{E}(\Delta W \, \Delta W |X_0) = \mathbf{I} \Delta t.
  \label{eq:wiener-process_properties}
\end{equation}
Increments over non-overlapping time intervals are statistically independent.
The stochastic term on the rhs of Eq.~\eqref{eq:white_noise} should be read using the so-called Ito-interpretation (see, e.g., \cite{Esp93, Gar97}).
This means that the expectation value of the increment is zero.
This requirement is consistent with the property of increments of fluctuating quantities as having a zero expectation value with respect to the initial state, eq.~\eqref{eq:projected_fluctuations_2}.

Note that in the expectation value in eq.~\eqref{eq:wiener-process_properties} is different than the expectation values in eq.~\eqref{eq:Langevin_integrated}.
In eq.~\eqref{eq:Langevin_integrated} the expectation value denotes an integration over all microstates $\Gamma$ that obey $X(\Gamma)=X_0$.
In eq.~\eqref{eq:wiener-process_properties} it is the expectation over the measure corresponding to the Wiener-process.
For this second expectation value, by means of Bayes theorem, $\mathbb{E}(\Delta W | X_0) = 0$ implies that $\mathbb{E}(X_0|\Delta W) = 0$.
This is automatically obeyed if we model $X_t$ to be an adapted (or non-anticipating) process.
This means that $X_t$ is independent of future events.
The assumption that $X_t$ is non-anticipating is always made when using stochastic differential equations.

Under the assumption of rapid decorrelating fluctuations, the generalized Langevin equation, eq.~\eqref{eq:langevin_X}, can be simplified to a stochastic differential equation.
First consider, eq.~\eqref{eq:langevin_X}, for $\Delta t \gg \tau$, then use the modeling assumption that $X_t$ is slow and that the fluctuating term can be modeled as Gaussian noise.
One obtains a stochastic difference equation, eq.~\eqref{eq:Langevin_integrated} (valid only after integration over $\Delta t \gg \tau$), which can be well approximated by the stochastic differential equation
\begin{equation}
  \begin{split}
  dX_t & = \mathbb{E}(\dot{X} | X_t) \, dt + \exp[-S] \\
  & \times \frac{\partial}{\partial X_{t}} \cdot \biggl [ \tilde{\ten{M}}^T \, \exp[S] \biggr ] \, dt + \sqrt{2 \ten{M}} \cdot dW_t \\ 
  & = \mathbb{E}(\dot{X} | X_t) \, dt + \tilde{\ten{M}} \cdot \frac{\partial S}{\partial X_{t}} \, dt + \frac{\partial}{\partial X_{t}} \cdot \tilde{\ten{M}}^T \, dt \\
  &\quad  + \sqrt{2 \ten{M}} \cdot dW_t.
  \end{split}
  \label{eq:SDE}
\end{equation}
This stochastic differential equation has three main contributions an instantaneous part, a biased part and a fluctuating (random) part.
The first term on the rhs gives the instantaneous change of $X_t$ averaged over all possible microstates consistent with this state.
The last term models the fluctuations with respect to this average motion.
On time scales larger than decorrelation time, $\tau$, this is effectively modeled by means of a white noise, or Wiener, process.
The biased part (at least the symmetric part) gives a drift toward macro states with higher entropy.
This bias can be explained intuitively by the argument that these regions correspond to a larger micro-phase-space-volume.

We will end this section with a philosophical note.
The approximation eq.~\eqref{eq:Langevin_integrated} is a controlled approximation of a formal result derived from reversible dynamics.
It is therefore valid irrespective of the direction of time.
When playing the movie of life backwards in time it remains (to good approximation) valid.
However, in this case approximating $\Delta X^\mathrm{fluct}_t$ as a stochastic process gives incorrect results.
Equation \eqref{eq:SDE} will predict that entropy increases (on average), but when playing the movie of life backwards entropy decreases.
We speculate that the fact that eq.~\eqref{eq:Langevin_integrated} can only be approximated well with a simple coarse-grained equation, such as e.g. eq.~\eqref{eq:SDE}, for integration in the future direction of time is an important part of the explanation of the arrow of time.

The behavior of humans (and animals) is based on predictions (what-if strategies) using prior information.
Our brain stores coarse-level scale information and uses this to plan future actions.
Making predictions, based on coarse-level information, i.e. $X$, is only possible in the ``forward direction'' of time.
Coarse-grained equations such as eq.~\eqref{eq:SDE} can give accurate predictions in the direction in which entropy increases (on average).
Therefore making predictions, based on coarse-level information, is only possible in the direction of increasing entropy.
Therefore human behavior, with the usual asymmetric notion of past and future, is only possible in the direction of increasing entropy.
The statement that entropy increases in the direction of time called future is a tautology.
 
\subsection{Change of variables}

As a preparation on the description of coarse-graining in the next section let us first consider the behavior of a change of variables for the derived stochastic differential equation.
This will illustrate a non-trivial transformation rule for the entropy.
Let us consider the injective (one-to-one) transformation $Y(X)$.

One way to obtain the governing equation for $Y$ is the use of Ito-calculus.
To most common way to write stochastic differential equations is the Ito-interpretation.
Here integrands in a time integral are evaluated at the initial time of each time-increment.
In this notation the Leibniz rule (chain rule) for differentiation is not valid.
The mean reason is the asymmetry between the treatment of the initial point and final point in a time-step.
A different notation is the Stratonovich interpretation.
Here integrands are, in a finite difference approximation, evaluated more symmetrically at point $X_0+ \tfrac{1}{2} \Delta X$ instead of $X_0$.
Heuristically, applying the chain-rule in the Stratonovich interpretation (indicated by an open dot) and applying a second order Taylor expansion one obtains the rules for Ito-calculus.
Here quadratic terms of $dW \, dW$ can be replaced by the expectation value $\ten{I} \, dt$, as given by, eq.~\eqref{eq:wiener-process_properties}.
These heuristic rules can, of course, be rigorously proved, \cite{Gar97}.
For transformation of stochastic differential equations the following rules hold,
\begin{widetext}
\begin{equation}
  \begin{split}
    dY &= \frac{\partial Y}{\partial X} \circ dX \equiv  \frac{\partial Y}{\partial X} \cdot dX + \frac{1}{2} \frac{\partial^2 Y}{\partial X^2} : dX dX = \frac{\partial Y}{\partial X} \cdot dX + \frac{\partial^2 Y}{\partial X^2}: \ten{M} \, dt\\
    & = \frac{\partial Y}{\partial X} \cdot \mathbb{E}(\dot{X} | X_t) \, dt + \exp[-S(X)] \frac{\partial Y}{\partial X} \cdot \frac{\partial}{\partial X} \cdot \Bigl [ \tilde{\ten{M}}^T \, \exp[S(X)] \Bigr ]  \, dt + \frac{\partial^2 Y}{\partial X^2}: \ten{M} \, dt + \frac{\partial Y}{\partial X} \cdot  \sqrt{2 \ten{M}} \cdot dW\\
    & = \mathbb{E}(\dot{Y} | Y) \, dt+ \exp[-S(Y)] \frac{\partial}{\partial Y} \cdot \Bigl [ ( \tilde{\ten{M}}^{y}) ^T \, \exp[S(Y)] \Bigr ]  \, dt +  \sqrt{2 \ten{M}^y} \cdot dW.
  \end{split}
  \label{eq:transformation-rule}
\end{equation}
\end{widetext}

The form of the equation of $Y$ should be the same as that of $X$ as is the case in last equation of eq.~\eqref{eq:transformation-rule}.
To obtain this last form the following transformation rules need to be used
\begin{equation}
  \begin{split}
    \mathbb{E}(\dot{Y} | Y) &= \frac{\partial Y}{\partial X} \cdot \mathbb{E}(\dot{X} | X),\\
    \tilde{\ten{M}}^y &= \frac{\partial Y}{\partial X} \cdot \tilde{\ten{M}} \cdot \frac{\partial Y}{\partial X},\\
    S(Y) &= S(X) - \ln\left|\det\Bigl(\frac{\partial Y}{\partial X}\Bigr)\right|.
  \end{split}
\end{equation}
The transformation of $\mathbb{E}(\dot{Y} | X)$ and $\tilde{M}$ follows immediately from the fact that one can take functions of $X$ (such as $\partial Y / \partial X$) out of the expectation value, eq.~\eqref{eq:cond-expect-prop}.
The transformation for the entropy might be more surprising.
Inserting the transformations into the last line of eq.~\eqref{eq:transformation-rule} and doing the required calculus proves the equalities.
Alternatively, it follows directly from the definition, eq.~\eqref{eq:entropy_definition},
\begin{multline}
    \exp[S(Y)] = \frac{dY_*(\mu_L)[Y]}{dY} = \frac{dX}{dY} \, \frac{dX_*(\mu_L)[X]}{dX} \\ 
    = \frac{\exp[S(X)]}{\left|\det\frac{\partial Y}{\partial X}\right|}.
    \label{eq:transformed_entropy}
\end{multline}
Here $dX/dY$ is the Radon-Nikodym derivative of the Lebesgue measure in the $X$ space to the Lebesgue measure in the $Y$ space, giving rise to the Jacobian determinant.
A somewhat more sloppy, but easier to handle, definition of the entropy is by means of the Dirac delta distribution,
\begin{multline}
  \exp[S(Y)] = \int \delta[Y-Y(\Gamma)] \, d\mu_L[\Gamma] \\ = \int \delta[Y-Y(X)] \, \exp[S(X)] \, dX
  = \frac{\exp[S(X)]}{\left|\det\frac{\partial Y}{\partial X}\right|}
\end{multline}
In this way the Jacobian determinant arises by means transformation of the delta distribution.

The final form in eq.~\eqref{eq:transformation-rule} is just the generic form eq.~\eqref{eq:SDE} (with $X$ replaced by $Y$).
The route via Ito-calculus is tedious.
Performing this exercise results in the important observation that to keep the equation in the correct form the non-scalar transformation rule of the entropy has to be taken into account.
Usually in thermodynamics one quickly goes to the thermodynamic limit (see \S\ref{sect:therm_limit}), and uses a scalar entropy.
Introducing a scalar entropy at the stage where fluctuations are important one runs into trouble.
If all terms in the equation would transform in a tensorial (or scalarian) way then the Ito-calculus would ruin the transformation.
The fact that the strange transforming entropy $S(X)$ appears in the equation saves the day.

\section{Successive coarse-graining}

In this section we will discuss how to coarse-grain an already coarse-grained description further.
Suppose the intermediate level is described by a state $X$, and the more coarse-grained level can be expressed as $Y(X)$.
Here the relation $X \rightarrow Y$ is many-to-one (surjective).
We will denote the (Zwanzig) projection operators as
\begin{equation}
  \mathcal{P} A = \mathbb{E}(A|X) \text{ and } \mathcal{P}^y A = \mathbb{E}(A|Y).
\end{equation}
Since $\mathbb{E}(\mathbb{E}(A|X)|Y)=\mathbb{E}(A|Y)$ one finds relations like $\mathcal{P}^y \mathcal{P} = \mathcal{P} \mathcal{P}^y = \mathcal{P}^y$ and $\mathcal{Q} \mathcal{Q}^y = \mathcal{Q}^y \mathcal{Q}= \mathcal{Q}$.
The decomposition of the coarse-grained dynamics of $Y_t$ obeys a similar equation as that of $X$  the difference is that here $\mathcal{P}^y$ instead of $\mathcal{P}$ is used for the decomposition.
In some of the steps it is more convenient to work with $\mathcal{P}^y$ instead of the conditional expectations.
So we will switch between the two representations.

The instantaneous part of the evolution equation of $Y$ can be obtained straight-forwardly,
\begin{equation}
  \begin{split}
    \mathcal{P}^y \mathcal{L} Y &= \mathbb{E}(\mathcal{L} Y|Y) = \mathbb{E}(\mathbb{E}(\mathcal{L} Y|X)|Y)\\
    &= E\Bigl(\mathbb{E}(\mathcal{L} X|X) \cdot \frac{\partial Y}{\partial X} \Bigl |Y \Bigl)\\ 
    &= \exp[-S(Y)] \, \int \mathbb{E}(\dot{X}|X) \cdot \frac{\partial Y}{\partial X} \\ & \quad \times \delta[Y - Y(X)] \, \exp[S(X)] \, dX.
  \end{split}
\end{equation}

To obtain the fluctuating contribution to the dynamics of $Y$ we want to evaluate $Y^{y,\mathrm{fluct}}_t \equiv \exp[\mathcal{Q}^y \mathcal{L} t] Y_0$.
For $Y_t^\mathrm{fluct}=\exp[\mathcal{Q} \mathcal{L} t] Y_0$ we have $Y_t^\mathrm{fluct} = Y(X_t^\mathrm{fluct})$.
This equality is a unique property of the Zwanzig formalism as given in eq.~\eqref{eq:coarse_grained_product_rule}.
Note, however, that $Y_t^{y,\mathrm{fluct}} \ne Y(X_t^{y,\mathrm{fluct}})$.
The reason is that $\mathcal{Q}^y \mathcal{L}$ does not act as a derivation (does not obey the chain rule) for functions of $X$.
Therefore, to construct $Y_t^{y,\mathrm{fluct}}$ one should, in the general case, consider both $Y_t^{y,\mathrm{fluct}}$ and $X_t^{y,\mathrm{fluct}}$.

When identifying
\begin{equation}
  \mathcal{Q}^y \mathcal{L} = \mathcal{P}\mathcal{Q}^y \mathcal{L} + \mathcal{Q} \mathcal{Q}^y\mathcal{L} = (\mathcal{P}-\mathcal{P}^y) \mathcal{L} + \mathcal{Q} \mathcal{L},
\end{equation}
this equality can be used to adapt, eq.~\eqref{eq:generalized_langevin_2}, with substitution $\mathcal{L} \rightarrow \mathcal{Q}^y \mathcal{L}$ to obtain for a general quantity $A$,
\begin{widetext}
  \begin{equation}
    \Delta A^{y,\mathrm{fluct}}_t =\int_0^t \exp[\mathcal{Q}^y \mathcal{L} {t'}]\, (\mathcal{P}-\mathcal{P}^y) \mathcal{L}  A_0 \, d{t'} + \int_0^t \exp[\mathcal{Q}^y \mathcal{L} {t'}]\, (\mathcal{P}-\mathcal{P}^y) \mathcal{L}  \Delta A^\mathrm{fluct}_{t-{t'}} \, d{t'}+ \Delta A^\mathrm{fluct}_t
  \end{equation}
  Similarly to the development of eq.~\eqref{eq:dissipative_term} one finds
  \begin{multline}
    (\mathcal{P} - \mathcal{P}^y) \mathcal{L}  \Delta A^\mathrm{fluct}_{t-{t'}} = \exp[-S(X)] \frac{\partial}{\partial X} \cdot \left( \exp[S(X)] \, \mathbb{E}(\dot{X}^\mathrm{fluct}_0 \, \Delta A^\mathrm{fluct}_{t-{t'}}|X)\right ) \\ -  \exp[-S(Y)] \frac{\partial}{\partial Y} \cdot \left( \exp[S(Y)] \, \mathbb{E}(\dot{Y}^{\mathrm{fluct}}_0 \, \Delta A^\mathrm{fluct}_{t-{t'}}|Y)\right ).
  \end{multline}
  In terms of conditional expectation values this gives,
  \begin{multline}
    \Delta A^{y,\mathrm{fluct}}_t =\int_0^t \biggl(E\bigl(\dot{A}_0|X^{y,\mathrm{fluct}}_{t'}\bigr) -E\bigl(\dot{A}_0|Y^{y,\mathrm{fluct}}_{t'}\bigr)\biggr) \, d{t'}\\
    + \int_0^t \Biggl( \exp[-S(X^{y,\mathrm{fluct}}_{t'})] \frac{\partial}{\partial X^{y,\mathrm{fluct}}_{t'}} \cdot \left( \exp[S(X^{y,\mathrm{fluct}}_{t'})] \, E\bigl(\dot{X}^\mathrm{fluct}_0 \, \Delta A^\mathrm{fluct}_{t-{t'}}|X^{y,\mathrm{fluct}}_{t'}\bigr)\right ) \\
    -  \exp[-S(Y^{y,\mathrm{fluct}}_{t'})] \frac{\partial}{\partial Y^{y,\mathrm{fluct}}_{t'}} \cdot \left( \exp[S(Y^{y,\mathrm{fluct}}_{t'})] \, E\bigl(\dot{Y}^{\mathrm{fluct}}_0 \, \Delta A^\mathrm{fluct}_{t-{t'}}|Y^{y,\mathrm{fluct}}_{t'}\bigr)\right ) \Biggr) \, d{t'}+ \Delta A^\mathrm{fluct}_t
    \label{eq:coarse-grained-fluctuations}
  \end{multline}
\end{widetext}
Since $X^\mathrm{fluct}_t$ and its statistics are known (note that $A^\mathrm{fluct}_t=A(X^\mathrm{fluct}_t)$) one can solve this equation, in principle.
Note that because (in general), $Y^{y,\mathrm{fluct}}_t\ne Y(X^{y,\mathrm{fluct}}_t)$, one needs to simultaneously solve these equations for $A^{y,\mathrm{fluct}}_t=X^{y,\mathrm{fluct}}_t$ and $A^{y,\mathrm{fluct}}_t=Y^{y,\mathrm{fluct}}_t$.

\subsection{The linear case}
\label{sec:linear}

To get a feeling for the general equation we will look at the special case where
\begin{equation}
  Y = \ten{B} \cdot X \text{ and } S(X) = - \frac{1}{2} X \cdot \ten{\Lambda} \cdot X \;
\end{equation}
which gives using the entropy definition that,
\begin{equation}
  S(Y)=c-\frac{1}{2} Y \cdot \ten{\Lambda}^y \cdot Y, \text{ with } \ten{\Lambda}^y=(\ten{B} \cdot \ten{\Lambda}^{-1} \cdot \ten{B})^{-1}. 
  \label{eq:linear-cg-entropy}
\end{equation}
Here $\ten{B}$ and $\ten{\Lambda}$ are taken to be independent of $X$.
Furthermore we will assume that $\tilde{\ten{M}}_t = \mathbb{E}(\Delta X^\mathrm{fluct}_t \dot{X}^\mathrm{fluct}_0|X)$ is also independent of $X$.
The instantaneous part instantaneous, $\mathbb{E}(\dot{X}|X)$, will be left out of consideration here and is taken to be zero.

Note that our starting point is, explicitly, not a stochastic differential equation.
Making that approximation would mean that we lose the information on $\dot{X}^{\mathrm{fluct}}_0$.
The derivation is much simpler if this information is still available.
Clearly, in practice, we often start from the stochastic level.
Some subtleties that arise in this case will be discussed in \S\ref{sect:Omega-A}.

The final goal is to find an expression for $\mathbb{E}(\Delta Y_t^{y,\mathrm{fluct}} \dot{Y}_0^{y,\mathrm{fluct}}|Y_s)$.
When this expression is known the coarse-grained equation for $Y$ can be written down.
In this special case one has $Y_t^{y,\mathrm{fluct}} = \ten{B} \cdot X_t^{y,\mathrm{fluct}}$.
Therefore we only need to study $X_t^{y,\mathrm{fluct}}$ and can obtain the information on $X_t^{y,\mathrm{fluct}}$ from this.
Inserting the assumptions into eq.~\eqref{eq:coarse-grained-fluctuations} gives
\begin{multline}
    \Delta X^{y,\mathrm{fluct}}_t = \\- \int_0^t \tilde{\ten{M}}_{t-t'}  \cdot \bigl (\ten{\Lambda} - \ten{B} \cdot \ten{\Lambda}^y \cdot \ten{B} \bigr) \cdot X^{y,\mathrm{fluct}}_{t'} \, dt' \\+ \Delta X^\mathrm{fluct}_t
    \label{eq:coarse-grained-fluctuations_2}
\end{multline}
This equation can be solved using a Laplace transform.
After performing this transform we multiply by $\dot{X}^{y,\mathrm{fluct}}_0 = \dot{X}^{\mathrm{fluct}}_0$ to obtain $\tilde{\ten{M}}_t$ and $\tilde{\ten{M}}^{y,x}_t = \mathbb{E}(\Delta{X}^{y,\mathrm{fluct}} \dot{X}^{y,\mathrm{fluct}}_0 |X)$.
this gives
\begin{equation}
  \begin{split}
    [1+\tilde{\ten{M}}_s \cdot \tilde{\ten{\Lambda}}] \cdot \tilde{\ten{M}}^{y,x}_s &= \tilde{\ten{M}}_s,\\
    \tilde{\ten{M}}^{y,x}_s &= [1+\tilde{\ten{M}}_s \cdot \tilde{\ten{\Lambda}}]^{-1} \cdot \tilde{\ten{M}}_s.
    \label{eq:Laplace-transform_M}
  \end{split}
\end{equation}
where $\tilde{\ten{\Lambda}} = (\ten{\Lambda} - \ten{B} \cdot \ten{\Lambda}^y \cdot \ten{B} \bigr)$.
Here the subscript $s$ indicates the Laplace-transform variable.
If one has a short correlation time $\tau$, then $\tilde{\ten{M}}_s = s^{-1} \tilde{\ten{M}}$, where $\tilde{\ten{M}}$ is independent of $s$, for $s \ll \tau^{-1}$.
For a stochastic differential equation the limit $\tau \rightarrow 0$ is the equality is found for all $s$.
For this limit we thus have
\begin{equation}
  \tilde{\ten{M}}^{y,x}_s = [s + \tilde{\ten{M}} \cdot \tilde{\ten{\Lambda}}]^{-1} \cdot \tilde{\ten{M}}.
  \label{eq:Laplace-transform_M_2}
\end{equation}
If $\lambda_k$ denote the eigenvalues of $\tilde{\ten{M}} \cdot \tilde{\ten{\Lambda}}$ this function of $s$ will be singular for values $s= -\lambda_k$.
Performing the inverse Laplace transform one obtains contributions that decay as $\exp[-\lambda_k \, t]$.
For times much larger than $\lambda_0^{-1}$ where $\lambda_0$ is the smallest non-zero eigenvalue these contributions have decayed.
 
One can construct a projection matrix from the left and right null-vectors of $\tilde{\ten{M}} \cdot \tilde{\ten{\Lambda}}$.
Therefore the right null space is determined by the null space of $\tilde{\Lambda}$ spanned by the ``columns'' of $\ten{\Lambda}^{-1} \cdot \ten{B}$, and the left one by the ``rows'' of $\ten{B} \cdot \ten{\Lambda}^{-1} \cdot\tilde{\ten{M}}^{-1}$.
This projection matrix is
\begin{multline}
    \ten{Q} = (\ten{\Lambda}^{-1} \cdot \ten{B} \cdot \ten{G}^{-1} \cdot \ten{B} \cdot \ten{\Lambda}^{-1} \cdot \tilde{\ten{M}}^{-1} )\\ \text{with } \ten{G} = \ten{B} \cdot \ten{\Lambda}^{-1} \cdot \tilde{\ten{M}}^{-1} \cdot \ten{\Lambda}^{-1} \cdot \ten{B}.
\end{multline}
It has the property that $\ten{Q} \cdot \tilde{\ten{M}} \cdot \tilde{\ten{\Lambda}} = \tilde{\ten{M}} \cdot \tilde{\ten{\Lambda}} \cdot \ten{Q} = 0$.
The matrix $\ten{G}$ is found from contraction of the vectors spanning the left and right null-spaces.

Applying this projection to eq.~\eqref{eq:Laplace-transform_M} one finds that
\begin{equation}
  \ten{Q} \cdot \tilde{\ten{M}}^{y,x}_s = s^{-1} \ten{Q} \cdot \tilde{\ten{M}},
\end{equation}
Applying $\ten{P} = 1-\ten{Q}$ to eq.~\eqref{eq:Laplace-transform_M_2} the contribution of $s^{-1} \ten{P} \cdot \tilde{\ten{M}} \cdot \tilde{\ten{\Lambda}} \cdot \ten{P}$ that decay quicker than $\lambda_0^{-1}$ remain, but the zero eigenvalues are filtered out.
This gives that for long times (when the other contributions have decayed) that
\begin{equation}
  s \, \tilde{\ten{M}}^{y,x}_s \rightarrow \ten{Q} \cdot \tilde{\ten{M}} \equiv \tilde{\ten{M}}^{y,x}.
\end{equation}
Using this expression we find that
\begin{multline}
  \tilde{\ten{M}}^y = \lim_{t \rightarrow \infty} \mathbb{E}(\Delta Y_t^{y,\mathrm{fluct}} \dot{Y}_0^{y,\mathrm{fluct}}|Y) \\
  = \ten{B} \cdot \ten{Q} \cdot \tilde{\ten{M}} \cdot \ten{B} = (\ten{\Lambda}^y)^{-1} \cdot \ten{G}^{-1} \cdot (\ten{\Lambda}^y)^{-1}.
  \label{eq:M_c-linear}
\end{multline}

Here we assumed that $\tilde{\ten{M}}$ and $\ten{\Lambda}$ are invertible.
This is not the most general case.
It is not even the typical case, because typically there are conserved quantities present in the system.
Usually the coarse-grained variables are chosen such that conserved quantities can be expressed using these variables.
These quantities give rise to null-vectors of $\tilde{\ten{M}}$ because conserved quantities do not fluctuate.
Also $\ten{\Lambda}$ can have zero eigenvalues.
For example, imagine the case where a Brownian particle is bound to the region near a plane by an entropic force, but is free to move in the direction parallel to the plane.
For the linear case, considered here, these unconstrained direction are not coarse-grained (otherwise $\ten{\Lambda}^y$ given by eq.~\eqref{eq:linear-cg-entropy} would be ill-defined).
For the singular matrix $\tilde{\ten{M}}$ the left null vectors are uninteresting.
They will give contributions to $\ten{Q}$ that when used in the multiplication $\ten{Q} \cdot \tilde{\ten{M}}$ will give zero.
For a singular matrix $\tilde{\ten{M}} \cdot \tilde{\ten{\Lambda}}$ we need to find vectors $\vec{v}_{\alpha}$ that are solution of the problem
We first construct the null-vectors of $\tilde{\ten{\Lambda}}$,
\begin{equation}
  \begin{split}
    \vec{u}_{\alpha} \cdot \ten{\Lambda} &= \vec{w}_{\alpha'} \cdot \ten{B}\\
    \vec{v}_{\alpha} \cdot \tilde{\ten{M}} &= \vec{u}_{\alpha}.  
  \end{split}
\end{equation}
Here the rhs in the last equation, i.e. $\vec{u}_{\alpha}$, needs to be in the null-space of $\tilde{\ten{\Lambda}}$, which gives rise to the first equation.
If $\ten{\Lambda}$ is singular there are more $\alpha$'s then $\alpha'$'s.
Furthermore $\vec{u}_{\alpha}$ should be in the range of $\tilde{\ten{M}}$ for a solution to exist. 
Let $\vec{z}_{\beta}$ be right null vectors of $\tilde{\ten{M}}$ (i.e. $\tilde{\ten{M}} \cdot \vec{z}_{\beta} = 0$) then
this means that $\vec{u}_{\alpha}$ has to obey
\begin{equation}
  \vec{u}_{\alpha} \cdot \vec{z}_{\beta}=0, \; \forall \beta.
\end{equation}
If $\tilde{\ten{M}}$ and $\ten{\Lambda}$ are invertible $\vec{w}_{\alpha}$ can be chosen to be the base vectors of the coarse-grained space and the previous result is recovered.

Clearly $\vec{v}_{\alpha}$ is determined up to a linear combination of left null-vectors of $\tilde{\ten{M}}$.
Therefore we will pose the extra condition
\begin{equation}
  \vec{v}_{\alpha} \cdot \vec{z}_{\beta} = 0, \;  \forall \beta.
\end{equation}
to fix $\vec{v}_{\alpha}$.
Having solved this equation one can define
\begin{equation}
  G_{\alpha\beta} = \vec{v}_{\alpha} \cdot \vec{u}_{\beta}, \; \ten{Q} =\vec{u}_{\alpha} [\ten{G}^{-1}]^{\alpha\beta} \vec{v}_{\beta}.
\end{equation}
The rank of $\ten{Q}$ is determined by the number of independent $\vec{w}_{\alpha}$'s.

Applying this projection to eq.~\eqref{eq:Laplace-transform_M} one finds that
\begin{equation}
    \tilde{\ten{M}}^y = \ten{B} \cdot \ten{Q} \cdot \tilde{\ten{M}} \cdot \ten{B}=\ten{B} \cdot \vec{u}_{\alpha} [\ten{G}^{-1}]^{\alpha\beta} \vec{u}_{\beta}\cdot \ten{B}
    \label{eq:M_c-linear2}
\end{equation}
Note that if $\tilde{\ten{M}}$ is symmetric (i.e. $\tilde{\ten{M}}=\ten{M}$) then also $\tilde{\ten{M}}^y$ is symmetric.
This can be seen from the fact that in this case $G_{\alpha\beta} = \vec{v}_{\alpha} \cdot \ten{M} \cdot \vec{v}_{\beta}$ is symmetric.

The general picture that arises is the following.
For the long time behavior there are ``constrained'' and ``unconstrained'' directions.
For directions into which $X^{y,fluct}$ changes but $Y^{y,fluct}$ not there is an restoring entropic driving force.
For directions of $X^{y,fluct}$ in the $Y^{y,fluct}$-plane there is unconstrained motion.
This motion is filtered out by $\ten{Q}$ and contributes to $\ten{M}^y$.
We believe that this picture remains valid for the general equation, eq.~\eqref{eq:coarse-grained-fluctuations}.

\subsection{Connection to homogenization theory} 

To illustrate the derived formulas we will give an outline for the case of diffusion.
In the current paper we will not go into detail of deriving continuum equations.
This example is only to illustrate the power of the derived relation.
We will consider the case of a spatial concentration that deviates a little bit from the equilibrium concentration.
The main variables are $\delta c(\vec{r}) = c(\vec{r}) - c_\mathrm{eq}(\vec{r})$.
By considering a small deviation we can remain in the (linear) framework outlined above.
The identification with the general theory is $X \rightarrow \delta c$.
The coordinates in space play the role of indexes, i.e., $X^i \rightarrow \delta c(\vec{r}_i)$.
The contractions indicated by the dots are replaced by integrals.
The entropy is given by
\begin{multline}
  S = - \int c(\vec{r}) \ln \left(\frac{c(\vec{r})}{c_\mathrm{eq}(\vec{r})}\right) \, d^3\vec{r} = \\S_0 - \int \frac{\delta c(\vec{r}) \, \delta c(\vec{r})}{c_\mathrm{eq}} \, d^3\vec{r}
\end{multline}
For this expression we can identify $\Lambda(\vec{r}, \vec{r}') = c_\mathrm{eq}^{-1}(\vec{r}) \delta(\vec{r}-\vec{r}')$.
The matrix $\tilde{\ten{M}}$ can be identified as
\begin{equation}
  f \cdot \tilde{\ten{M}} \cdot g = \int \nabla f(\vec{r}) \cdot c_\mathrm{eq}(\vec{r}) \ten{D}(\vec{r}) \cdot \nabla g(\vec{r}) \, d^3\vec{r}.
\end{equation}
Here $\ten{D}(\vec{r})$ is the (position dependent) diffusion matrix.
This ``matrix'' is symmetric (so $\tilde{\ten{M}} = \ten{M}$), but also singular.
Constant functions, i.e. independent of $\vec{r}$, span the (1-dimensional) null-space of $\ten{M}$.

As coarse-grained variables let us consider a finite number of Fourier-modes $c_k$ corresponding to small wave vectors $\vec{k}$.
\begin{equation}
  \delta c_k = \int \exp[i \vec{k} \cdot \vec{r}]\, \delta c(\vec{r}) \, d^3\vec{r}, \text{ so } B^{j'}_{j} \rightarrow \exp[i \vec{k} \cdot \vec{r}]. 
\end{equation}
Therefore the null-vectors of $\tilde{\ten{\Lambda}}$, which give the right null vectors of $\tilde{\ten{M}} \cdot \tilde{\ten{\Lambda}}$ are $c_\mathrm{eq}(\vec{r}) \, \exp[i \vec{k} \cdot \vec{r}]$.
The right null vectors can be found by solving,
\begin{multline}
  - \nabla \cdot \Bigl (c_\mathrm{eq}(\vec{r}) \ten{D}(\vec{r}) \cdot \nabla v(\vec{k}, \vec{r}) \Bigr) = \\
    c_\mathrm{eq}(\vec{r}) \, \exp[i \vec{k}\cdot \vec{r}] - a(\vec{k}) \, c_\mathrm{eq}(\vec{r}) \sum_{\vec{k}'} \exp[i \vec{k}'\cdot \vec{r}].
\end{multline}
Here $a(\vec{k})$ should be chosen such that the right-hand side is in the range of $\ten{M}$,
\begin{equation}
  \begin{split}
    a(\vec{k}) &= \frac{c_{\mathrm{eq},k}}{\sum_{\vec{k}'} c_{\mathrm{eq},k'}}, \text{ with}\\
    c_{\mathrm{eq},k} &= \int c_\mathrm{eq}(\vec{r}) \exp[i \vec{k}\cdot \vec{r}] \, d^3\vec{r}.
  \end{split}
\end{equation}

If we assume that $c_\mathrm{eq}(\vec{r})$ does vary rapidly on small length scales but is homogeneous on larger length scales then $c_{\mathrm{eq},k}=0$ if the $\vec{k}$'s are small enough.
Solving the ``cell problem'' one finds $v(\vec{k}, \vec{r})$.
The effective diffusion coefficient is then given by:
\begin{equation}
  D^{-1}_\mathrm{eff}=\frac{k^2}{V \langle c_\mathrm{eq} \rangle} \int \vec{v}(\vec{r},\vec{k}) c_\mathrm{eq}(\vec{r}) \, \exp[-i \vec{k}\cdot \vec{r}] d^3\vec{r}.
\end{equation}
For the one dimensional case one finds that
\begin{equation}
  \begin{split}
      c_\mathrm{eq}(x) D(x) \frac{d}{dx} v(k, x) &\approx \sum_{x_j\le x} \int_{x_j}^{x_{j+1}} c_\mathrm{eq}(\vec{x'}) \, \exp[i k x'] dx' \\
      &\approx  \langle c_\mathrm{eq} \rangle \sum_{x_j\le x}  \exp[i k x'] \Delta x'\\
      &\approx -\frac{1}{ik} \langle c_\mathrm{eq} \rangle \exp[i k x]
  \end{split}
\end{equation}
Applying this procedure a second time then gives
\begin{equation}
  \begin{split}
      v(x,k) &\approx -\frac{1}{ik} \langle c_\mathrm{eq} \rangle \sum_{x_j\le x} \int_{x_j}^{x_{j+1}} [c_\mathrm{eq}(x') D(x')]^{-1} \, \exp[i k x'] dx'\\
      &\approx \frac{1}{k^2} \langle c_\mathrm{eq} \rangle \langle [c_\mathrm{eq} \, D]^{-1} \rangle \, \exp[i k x]\\
      D^{-1}_\mathrm{eff} &= \langle c_\mathrm{eq} \rangle \langle [c_\mathrm{eq} \, D]^{-1} \rangle.
  \end{split}
\end{equation}
The equilibrium density follows, e.g, from a rough background potential, $c_\mathrm{eq}(x) \propto \exp[-U(x)/kT ]$, the result predicts that diffusion is much hampered if potential energy differences a few times $kT$ are present.
For the case of constant $D$ the equation is a classical result, see e.g. \cite{Zwa88}.

The general result in more dimensions can also be found be means of homogenization techniques \cite{Pap95,Cio99}.
The presented procedure is, however, more general.
It gives a general recipe for (near-equilibrium) coarse graining.
It is valid when properties on the fine scales are very rough.
It also gives the recipe of how to treat a coarse graining if there is no wide separation of scales, such that homogenization techniques are not valid.

By means of eq.~\eqref{eq:Laplace-transform_M_2} it also indicates the range of validity of the obtained result.
There is a restriction on time-scales.
Coarse-grained equations are only useful to study phenomena above a certain length-scale.
The non-zero eigenvalues of $\tilde{\ten{M}} \cdot \tilde{\ten{\Lambda}}$ are decay rates that appear by means of poles in the inverse Laplace transform.
If the degree of coarse-graining is very large the spectrum of eigenvalues is (almost) continuous.
In this case one might find, e.g., power-law region.

\subsection{The instantaneous part}

The instantaneous (ensemble averaged) rate of change, i.e., $\mathbb{E}(\dot{X}|X)$ inherits the phase space incompressible property as given by eq.~\eqref{eq:degeneracy}.
This gives, that under certain restrictions concerning topology and smoothness, 
\begin{equation}
  \mathbb{E}(\dot{X}|X) = \exp[-S(X)] \frac{\partial}{\partial X} \cdot \Bigl( \ten{\Omega}^T \exp[S(X)] \Bigr ),
  \label{eq:instantaneous}
\end{equation}
where $\ten{\Omega} = - \ten{\Omega}^T$.
This is a consequence of Stokes theorem.
Upon coarse-graining this matrix transforms as
\begin{equation}
  \ten{\Omega}^y = E \biggl(\frac{\partial Y}{\partial X} \cdot \ten{\Omega} \cdot \frac{\partial Y}{\partial X} \biggl |Y \biggr).
  \label{eq:coarse-graining_Omega}
\end{equation}
This can be checked by evaluating the coarse-graining from $\mathbb{E}(\dot{X}|X)$ to $\mathbb{E}(\dot{Y}|Y)$ and inserting eq.~\eqref{eq:instantaneous}.
So if one can find $\ten{\Omega}$ at one level one can find an expression on any level (and upon a change of variables).

At the microscopic level Hamilton dynamics holds.
In the canonical form this can be written as
\begin{equation}
  \dot{\Gamma} = \frac{\partial}{\partial \Gamma} \cdot \bigl ( \ten{L}_\mathrm{micro}^T \, H(\Gamma) \bigr). 
\end{equation}
Here $\ten{L}_\mathrm{micro}$ is a constant anti-symmetric matrix.
It can be written in a block-diagonal form , where the $2\times 2$ blocks have $\pm 1$ as off-diagonal elements.
Since, at this level of description $S(\Gamma)=0$, this is of the required form.
Coarse-graining the microscopic form gives
\begin{equation}
  \ten{\Omega} = E\biggl(\frac{\partial X}{\partial \Gamma} \cdot \ten{L}_\mathrm{micro}  \cdot \frac{\partial X}{\partial \Gamma} \, H(\Gamma) \biggl |X \biggr).
  \label{eq:Omega_def}
\end{equation}
Note that, because total energy is a conserved quantity, one often chooses the coarse-grained variables such that the total energy can be expressed as function of $X$.
In that case $H(\Gamma)=H(X(\Gamma))$ and the energy can be taken out of the expectation value, such that
\begin{equation}
  \ten{\Omega} = \ten{L} H(X), \text{ where } \ten{L} = E\biggl(\frac{\partial X}{\partial \Gamma} \cdot \ten{L}_\mathrm{micro}  \cdot \frac{\partial X}{\partial \Gamma} \biggl |X \biggr)
  \label{eq:L-matrix}
\end{equation}
For the quantity $\ten{L}$ one finds the degeneracy condition, similar to the one of $\mathbb{E}(\dot{X}|X)$, eq.~\eqref{eq:degeneracy},
\begin{equation}
  \exp[-S(X)] \frac{\partial}{\partial X} \cdot \bigl ( \ten{L}^T \exp[S(X)] \bigl) =0.
  \label{eq:degeneracy_L}
\end{equation}
The essential step in the proof is that
\begin{multline}
  \frac{\partial}{\partial X} \cdot \int \frac{\partial X}{\partial \Gamma} \cdot \ten{L}_\mathrm{micro}^T  \cdot \frac{\partial X}{\partial \Gamma} \delta[X-X(\Gamma)] d\mu_L[\Gamma] \\ = \int \frac{\partial X}{\partial \Gamma} \cdot \ten{L}_\mathrm{micro} \cdot \frac{\partial X}{\partial \Gamma} \cdot \frac{\partial}{\partial X} \delta[X-X(\Gamma)] d\mu_L[\Gamma]\\= - \int \frac{\partial X}{\partial \Gamma} \cdot \ten{L}_\mathrm{micro} \cdot \frac{\partial}{\partial \Gamma} \delta[X-X(\Gamma)] d\mu_L[\Gamma].
\end{multline}
The final steps consist of partial integration and using the fact that $\ten{L}_\mathrm{micro}$ is constant and anti-symmetric.
Applying the degeneracy condition eq.~\eqref{eq:degeneracy_L}, one can write
\begin{equation}
  \mathbb{E}(\dot{X}|X) = \ten{L} \cdot \frac{\partial H}{\partial X}.
  \label{eq:instantaneous_Hamltonian}
\end{equation}
Note that, when energy can be expressed in terms of $X$ then $H(X)$ can not fluctuate.
This results into
\begin{equation}
  \frac{\partial H}{\partial X} \cdot \tilde{\ten{M}} = 0, \; \tilde{\ten{M}} \cdot \frac{\partial H}{\partial X}=0.
  \label{eq:degeneracy_M}
\end{equation}
Conditions eq.~\eqref{eq:degeneracy_L} and eq.~\eqref{eq:degeneracy_M} are the degeneracy conditions of the GENERIC formalism.
In the original formulation a simpler degeneracy condition was given for $\ten{L}$ was only valid in the thermodynamic limit, \cite{Ott97_2}.
This was corrected in \cite[eq.~{(6.163)}]{Ott05}.
We were not able to proof the GENERIC claim that the instantaneous part is necessarily of a Poisson brackets form (obeying the Jacobi identity).

Let's repeat the exercise from the previous section, i.e. coarse-graining in the linear case, including a constant $\ten{\Omega}$.
Coarse-graining the instantaneous part is straightforward,
\begin{equation}
  \ten{\Omega}^y=\ten{B} \cdot \ten{\Omega} \cdot \ten{B}.
\end{equation}
For determining $\tilde{\ten{M}}^y$ a term $-\int_0^t \ten{\Omega} \cdot \tilde{\ten{\Lambda}} \cdot X_{t'} dt'$ needs to be added to eq.~\eqref{eq:coarse-grained-fluctuations_2}.
One consequence is that
\begin{equation}
  \dot{X}^{y,\mathrm{fluct}}_0 = \ten{\Omega} \cdot \tilde{\ten{\Lambda}} \cdot X_0 - \dot{X}^{\mathrm{fluct}}_0.
\end{equation}
This has no influence on the derivation since $\mathbb{E}(X_0 \Delta X^{\mathrm{fluct}}_t|X_0) = 0$.
The final equation we find is
\begin{equation}
    \tilde{\ten{M}}^{y,x}_s = [1+(s^{-1} \ten{\Omega} + \tilde{\ten{M}}_s) \cdot \tilde{\ten{\Lambda}}]^{-1} \cdot  \tilde{\ten{M}}_s.
  \label{eq:linear_coarse-graining_omega}
\end{equation}
In principle the procedure for finding $\ten{M}^y$ is the same as the method presented in \S\ref{sec:linear}.

If we assume that the coarse-graining is such that the energy can be expressed as a function of the coarse-grained variables (e.g., because internal energy is a coarse-grained variable), this means that $H(X)=H(Y(X))$.
In this case the instantaneous part can be written as,
\begin{equation}
  \begin{split}
    \mathbb{E}(\dot{A}|X) &= \frac{\partial A}{\partial X} \cdot \ten{L} \cdot \frac{\partial H}{\partial X} = \frac{\partial A}{\partial X} \cdot \ten{L} \cdot \frac{\partial Y}{\partial X} \cdot \frac{\partial H}{\partial Y}\\
    \mathbb{E}(\dot{A}|Y) &= E\Bigl (\frac{\partial A}{\partial X} \cdot \ten{L} \cdot \frac{\partial Y}{\partial X} \Bigr |Y \Bigr) \cdot \frac{\partial H}{\partial Y}
  \end{split}
\end{equation}
Therefore the instantaneous contribution to eq.~\eqref{eq:coarse-grained-fluctuations}, becomes
\begin{multline}
  \mathbb{E}(\dot{A}|X) - \mathbb{E}(\dot{A}|Y) = \\
  \left( \frac{\partial A}{\partial X} \cdot \ten{L} \cdot \frac{\partial Y}{\partial X}  - E\Bigl (\frac{\partial A}{\partial X} \cdot \ten{L} \cdot \frac{\partial Y}{\partial X} \Bigr |Y \Bigr) \right) \cdot \frac{\partial H}{\partial Y}.
\end{multline}
Note that, for the case of constant $\ten{L}$ and $\ten{B}$ the term $\mathbb{E}(\dot{X}|X) - \mathbb{E}(\dot{X}|Y)$ equals zero.
Therefore, for this shape of instantaneous part, the coarse-graining does not contribute to $X^{y,\mathrm{fluct}}$,  so eq.~\eqref{eq:M_c-linear2} is not influenced.
This means that for a (non-constant) $\ten{\Omega}$ of the form,
\begin{equation}
  \ten{\Omega}(X) = \ten{L} \, H(Y(X)) + \tilde{\ten{\Omega}},
\end{equation}
where $\ten{L}$ and $\tilde{\ten{\Omega}}$ are constant, one finds eq.~\eqref{eq:linear_coarse-graining_omega} with $\ten{\Omega}$ replaced by $\tilde{\ten{\Omega}}$.

\subsection{Onsager-Casimir symmetries}

Let's investigate the symmetric and antisymmetric parts of $\ten{A}$ somewhat further.
Usually $\tilde{\ten{M}}$ is taken to be symmetric because one expects it to obey the Onsager relations \cite{Gro62}.
Casimir showed that in special cases there can also be an anti-symmetric contribution $\ten{A}$. 
We will investigate these claims in our framework, also out of equilibrium.

The reasoning is as follows.
Microscopic dynamics are reversible.
This means that to any micro state $\Gamma$ a time reversed state can be associated.
Let's denote $\mathcal{T}$ as the time reversal operator, then for any $t$,
\begin{equation}
  \exp[-\mathcal{L}t] = \mathcal{T} \, \exp[\mathcal{L}t] \, \mathcal{T},
\end{equation}
(and from this $\mathcal{T}^2=1$ and $\mathcal{T} \mathcal{L} + \mathcal{L}\mathcal{T}=0$).
For a micro state $\Gamma$ there is a one-to-one functional relation $\mathcal{T} \Gamma = T(\Gamma)$.
Using the fact that the microscopic Liouville operator acts as a derivation one finds that for the phase space velocity
\begin{equation}
  \mathcal{T} \dot{\Gamma} = - \dot{\Gamma} \cdot \frac{\partial T}{\partial \Gamma},
\end{equation}
and since $\mathcal{T}^2=1$,
\begin{equation}
  \frac{\partial T}{\partial \Gamma} \cdot \frac{\partial T}{\partial \Gamma} = 1, \text{ so that } \det \left( \frac{\partial T}{\partial \Gamma} \right) = \pm 1.
\end{equation}
Usually a time-reversal operation corresponds to $\mathcal{T} \vec{r}^i = \vec{r}^i$ and a change of sign of the momenta, $\mathcal{T} \vec{p}_i = - \vec{p}_i$.

Now one assumes that the coarse-graining is performed such that upon time-reversal, also for the coarse-grained space, there is a functional relation $\mathcal{T} X = T^x(X)$.
Usually this established by making the members of $X$ only to depend on even or odd powers of the momenta, such that $\mathcal{T} X^i =\pm X^i$.
One consequence of this definition is that
\begin{equation}
  \exp[S(T^x(X))] = \frac{\exp[S(X)]}{\left|\det\frac{\partial T^x}{\partial X}\right|} = \exp[S(X)],
\end{equation}
because, for the same reasons as in the microscopic case, the determinant is $\pm 1$.
For the expectation values one finds that
\begin{equation}
  \begin{split}
    \mathcal{T} \mathbb{E}(A|X) &= \mathbb{E}(A|T^x(X))\\
    &= \exp[-S(X)] \int A(\Gamma) \, \delta[X(\Gamma) - T^x(X) ] \, d\mu_L[\Gamma]\\
    &= \exp[-S(X)] \int A(T(\Gamma)) \, \delta[X(\Gamma) - X ] \, d\mu_L[\Gamma]\\
    &= \mathbb{E}(\mathcal{T} A|X).    
  \end{split}
\end{equation}
Here we extensively used the fact that $\mathcal{T}$ does leave measures (and therefore also the delta distribution) invariant.
In terms of projection operators we thus have proved that, $\mathcal{T}$ and $\mathcal{P}$ do commute,
\begin{equation}
  \mathcal{T} \mathcal{P} - \mathcal{P} \mathcal{T}=0.
\end{equation}
Therefore one can pull $\mathcal{T}$ through projection operators such that, e.g., 
\begin{equation}
  \mathcal{T} \exp[\mathcal{Q} \mathcal{L} t] = \exp[- \mathcal{Q} \mathcal{L} t] \mathcal{T}.
\end{equation}

For the instantaneous part we find that upon time-reversal
\begin{equation}
  \begin{split}
    \mathcal{T} \mathbb{E}(\dot{X}|X) &= \mathbb{E}(\mathcal{T} \mathcal{L} X|X)= - \mathbb{E}(\mathcal{L} \mathcal{T} X|X) \\
    &= - \mathbb{E}(\mathcal{L} \, T^x(X)|X)= - \frac{\partial T^x}{\partial X} \cdot \mathbb{E}(\dot{X}|X). 
  \end{split}
\end{equation}
Which also gives that
\begin{equation}
  \ten{\Omega}(T^x(X)) = \mathcal{T} \ten{\Omega} = - \frac{\partial T^x}{\partial X} \cdot \ten{\Omega}(X) \cdot \frac{\partial T^x}{\partial X}.
  \label{eq:onsager-casimir_Omega}
\end{equation}

For the correlations of the fluctuating contributions one finds that
\begin{multline}
    \mathcal{T} \mathbb{E}( \dot{X}^\mathrm{fluct}_0 \, \dot{X}^\mathrm{fluct}_\tau | X) =\\
     \frac{\partial T^x}{\partial X} \cdot \mathbb{E}( \dot{X}^\mathrm{fluct}_0 \, \dot{T^x}(X^\mathrm{fluct}_{-\tau}) | X).
\end{multline}
Here we used to following relations,
\begin{equation}
  \begin{split}
    \mathcal{T} \dot{X}^\mathrm{fluct}_\tau &= \mathcal{T} \mathcal{Q} \mathcal{L} \, \exp[\mathcal{Q} \mathcal{L} \tau] X_0\\
    & = - \mathcal{Q} \mathcal{L} \, \exp[\mathcal{Q} \mathcal{L} \tau] T^x(X_0)\\
    &= - \mathcal{Q} \mathcal{L} T^x(X^\mathrm{fluct}_{-\tau})\\
    &= - \dot{T}^x(X^\mathrm{fluct}_{-\tau}).
  \end{split}
\end{equation}
Using a similar approach as at eq.~\eqref{eq:symmetric_M}, i.e. assuming that during a typical decorrelation of the fluctuating contribution the change of $X$ is small, one finds that
\begin{equation}
  \tilde{\ten{M}}(T^x(X)) = \frac{\partial T^x}{\partial X} \cdot \tilde{\ten{M}}^T(X) \cdot \frac{\partial T^x}{\partial X}.
  \label{eq:Onsager-Casimir_M}
\end{equation}

The matrix $\partial T^x / \partial X$ can always be diagonalized with $\pm 1$ on the diagonals.
These eigenvalues indicate the parities upon time reversal.
Usually the natural choice of variables is such that the matrix has this diagonal form.

Let's consider a constant matrix $\tilde{\ten{M}}$, i.e., independent of $X$.
If the diagonalization is performed such that  $+1$'s are collected on the upper diagonal and the $-1$'s on the lower one, the $\tilde{M}$-matrix (in the basis provided by the eigenvectors) will have the form
\begin{equation}
\tilde{\ten{M}}=
\left( \begin{matrix}
  \ten{M}^{11} & \ten{A}^{12}\\
  - (\ten{A}^{12})^T & \ten{M}^{22}
\end{matrix}
\right )
=\left( \begin{matrix}
  (\ten{M}^{11})^T & -(\ten{A}^{12})^T\\
  \ten{A}^{12} & (\ten{M}^{22})^T
\end{matrix}
\right ).
\end{equation}
The symmetric part of $\tilde{\ten{M}}$ is due to quantities that have the same parity upon time-reversal.
The anti-symmetric part is due to the interaction of quantities with opposite parity.

Since, by definition, $\ten{\Omega}$ is anti-symmetric one gets from eq.~\eqref{eq:onsager-casimir_Omega} for a constant $\ten{\Omega}$ matrix
\begin{equation}
\ten{\Omega}=
\left( \begin{matrix}
  0& \ten{\Omega}^{12}\\
  - (\ten{\Omega}^{12})^T & 0
\end{matrix}
\right )=
\left( \begin{matrix}
  0& -(\ten{\Omega}^{12})^T\\
  \ten{\Omega}^{12} & 0
\end{matrix}
\right )
.
\end{equation}
So, also with respect to the Onsager-Casimir symmetry $\ten{A}$ and $\ten{\Omega}$ behave the same.
Lastly, because the entropy is invariant under time-reversal, we find that for $\ten{\Lambda}$
\begin{equation}
  \ten{\Lambda}=
  \left( \begin{matrix}
  \ten{\Lambda}_{11} & 0\\
  0 & \ten{\Lambda}_{22}
\end{matrix}
  \right ).
\end{equation}

Note that the Onsager-Casimir symmetries are strictly valid only for constant matrices $\tilde{\ten{M}}$ and $\ten{\Omega}$.
If the matrices are $X$ dependent then the derived relation, eq.~\eqref{eq:Onsager-Casimir_M}, relates the matrix at $X$ with the matrix at $T^x(X)$.
If this relation is simple, say entries of $\tilde{\ten{M}}(T^x(X))$ are $\pm$ entries of $\tilde{\ten{M}}(X)$ one can derive generalized Onsager-Casimir relations.
In \cite{Ott05} these relations are called ``dressed'' Onsager-Casimir symmetries.

\subsection{Instantaneous, reversible, isentropic}
\label{sect:Omega-A}

In our theory, at the level of the stochastic differential equation, we have three matrices $\ten{\Omega}$, $\ten{A}$ and $\ten{M}$.
The matrix $\ten{\Omega}$ characterizes instantaneous response of the system (averaged over all microstates $X(\Gamma)=X$).
Upon coarse-graining it transforms according to eq.~\eqref{eq:coarse-graining_Omega}.

The matrices $\ten{A}$ and $\ten{M}$ follow from the time-correlation of the fluctuations.
They are non-instantaneous.
For the special case where all terms are linear eq.~\eqref{eq:linear_coarse-graining_omega} gives the auto-correlation of fluctuations upon coarse-graining.
The result is given in terms of a Laplace transform which can be used to obtain the coarse-grained value $\ten{M}$.
Upon coarse graining of $\ten{A}$ and $\ten{M}$ time-correlations enter. 
The use of a stochastic differential equation for the further coarse-grained equation is only a good approximation of the decorrelation time corresponding to these fluctuations is small enough.

Sometimes the instantaneous term, corresponding to $\ten{\Omega}$, is called the reversible term.
We think that the combined term $\ten{\Omega} + \ten{A}$ deserves this name.
Both terms are anti-symmetric.
From the Onsager-Casimir symmetries both terms can be identified as reversible.
If we write
\begin{equation}
  \dot{X}^\mathrm{rev} = \exp[-S(X)] \frac{\partial}{\partial X} \cdot \Bigl( (\ten{\Omega}^T + \ten{A}^T) \exp[S(X)] \Bigr ),
\end{equation}
this reversible contribution obeys
\begin{equation}
  \exp[-S(X)] \frac{\partial}{\partial X} \cdot \bigl( \dot{X}^\mathrm{rev} \, \exp[S(X)] \bigr ) =0,
  \label{eq:isentropic}
\end{equation}
as do the terms individually.
This can be seen as a generalization of the Liouville-theorem.
It is also a generalization of the isentropic evolution of reversible motion.
In fact, as we will discuss below, the isentropic condition follows from this when the thermodynamic limit is valid.

If one splits $\ten{\Omega}$, as 
\begin{equation}
  \ten{\Omega} = \tilde{\ten{\Omega}} + \ten{L} \cdot H(X),
\end{equation}
then
\begin{multline}
  \dot{X}^\mathrm{rev} = \ten{L} \cdot \frac{\partial}{\partial X} H \\+ \exp[-S(X)] \frac{\partial}{\partial X} \cdot \Bigl( (\tilde{\ten{\Omega}}^T + \ten{A}^T) \exp[S(X)] \Bigr )
\end{multline}
For this form eq.~\eqref{eq:isentropic} can be seen to hold by invoking eq.~\eqref{eq:degeneracy_L} and the anti-symmetry of $\ten{L}$.

The remaining term $\ten{M}$ gives rise to irreversible motion (and fluctuations).
\begin{equation}
  dX^\mathrm{irr} = \exp[-S(X)] \frac{\partial}{\partial X} \cdot \Bigl(\ten{M} \exp[S(X)] \Bigr ) \, dt + \sqrt{2 \ten{M}} \cdot d\vec{W},
\end{equation}
The matrix $\ten{M}$ is semi-definite symmetric.
Terms relating quantities of opposite parity are zero.
In the thermodynamic limit this term reduces to a dissipative term $\ten{M}\cdot {\partial S}/{\partial X}$ that always has a positive entropy production.
For smaller systems there can be negative entropy production (but not on average).

The full stochastic differential equation then becomes
\begin{equation}
  dX_t = \dot{X}^\mathrm{rev} \, dt + dX^\mathrm{irr}.
\end{equation}
Note that there is a very curious situation now.
In this equation $\ten{\Omega}+\ten{A}$, or at least $\tilde{\ten{\Omega}}+\ten{A}$ appears as one term.
Both $\tilde{\ten{\Omega}}$ and $\ten{A}$ seem to have similar properties.
If one however looks at expressions of a more coarse-grained situation, say $\tilde{\ten{\Omega}}^y$ and $\ten{A}^y$, the matrices $\tilde{\ten{\Omega}}$ and $\ten{A}$ enter in a different way.
One would hope that the expression for $\tilde{\ten{\Omega}}^y + \ten{A}^y$ would depend on the sum $\tilde{\ten{\Omega}}+\ten{A}$, but not on $\tilde{\ten{\Omega}}$ and $\ten{A}$, individually.
We were not able to proof this.

We thus have thus the curious result that a class of stochastic differential equation for $Y$ are all consistent coarse-grainings the same stochastic differential equation for $X$, but correspond to different microscopic dynamics.
This situation is concerned with the fact that in a stochastic differential equation $\dot{X}$ is undetermined.
Therefore one is unable to uniquely determine the anti-symmetric part of $\tilde{\ten{M}}^y$.
This was one of the reasons that in \S\ref{sec:linear} we did not take a stochastic differential equation as starting point.
In many situations one puts $\ten{A}=0$ from the beginning.
However, upon coarse-graining, a non-symmetric contribution can pop-up from the contributions of the instantaneous part (see eq.~\eqref{eq:linear_coarse-graining_omega}).
The situation remains a bit unsatisfactory from a conceptual point of view.

\section{Non-equilibrium thermodynamics}

The goal of non-equilibrium thermodynamics is to supply a description of the time-evolution of a system in terms of coarse-grained, meso- or macroscopic, variables.
The generalized non-linear Langevin equation, after approximation for the fluctuating forces, supplies such a description.

Therefore the derived equations provide a description that can be called non-equilibrium thermodynamics.
The theory deserves this predicate because entropy appears in it, and plays a central role.
The entropy that appears in the theory is a microcanonical entropy.
This is the basic definition.
It is defined as (the logarithm of) a density of states.
Therefore, upon coordinate transformation, eq.~\eqref{eq:transformed_entropy}, it does not transform as a scalar.
One might object this is not the way entropy should behave.
One might think that entropy should be a scalar.
Besides the constructive derivation, we have shown, however, that this is exactly how entropy should behave to preserve the general form of the equations when fluctuations are present.

\subsection{The thermodynamic limit}
\label{sect:therm_limit}

In the thermodynamic limit entropy is expected to behave as a scalar.
The thermodynamic limit behavior of entropy, starting from the microcanonical entropy definition, is well understood using large-deviation theory \cite{Rue69, Lan73, Ell99}.

We will here give an outline of a personal interpretation of these results.
The central quantity in statistical mechanics is the sum of states.
The (generalized) sum of states is just the Laplace transform of the density of states,
\begin{equation}
  \begin{split}
    Z(\lambda) &= \int \exp[S(X)] \exp[-\lambda \cdot X] \, dX \\
    &= \int \int \delta[X-X(\Gamma)] \exp[-\lambda \cdot X] \, dX \, d\mu_L[\Gamma]\\ 
    &= \int \exp[-\lambda \cdot X(\Gamma)] \, d\mu_L[\Gamma].    
  \end{split}
  \label{eq:Laplace-transform_S}
\end{equation}
The (generalized) grand-potential is then defined as
\begin{equation}
  \Phi(\lambda) = -\ln Z(\lambda). 
\end{equation}
Note that, by definition,
\begin{multline}
  - \frac{\partial^2 \Phi(\lambda)}{\partial \lambda \partial \lambda} = \frac{1}{Z(\lambda)} \times \\ \int \Bigl (X(\Gamma) - \bar{X}\Bigr) \, \Bigl (X(\Gamma) - \bar{X}\Bigr) \, \exp[-\lambda \cdot X(\Gamma)] \, d\mu_L[\Gamma],
\end{multline}
is positive (semi)-definite for real $\lambda$.
This means that $\Phi(\lambda)$ is a concave function of $\lambda$.
In the formula we used the definition
\begin{equation}
  \bar{X} = \frac{\partial \Phi(\lambda)}{\partial \lambda}.
  \label{eq:equilibrium_X}
\end{equation}

Large systems can be decomposed out of $N$ more or less independent, equivalent, sub-systems.
Let us consider an extensive quantity, $X$, that is the sum of values $X_\mathrm{sub}$ attained in these sub-systems.  
\begin{equation}
  X(\Gamma) = \sum_\alpha X_\mathrm{sub}(\Gamma^\alpha)
\end{equation}
 one has,
\begin{equation}
  \begin{split}
    Z_N(\lambda) &= \int \exp[-\lambda \cdot \sum_{\alpha=1}^N X_\mathrm{sub}(\Gamma^\alpha)] \, \prod_{\alpha=1}^N d\mu_L[\Gamma^\alpha]\\
    &= \bigl ( Z_\mathrm{sub}(\lambda) \bigr )^N.
  \end{split}  
\end{equation}
If the $N$ subsystems are equivalent one thus finds that
\begin{equation}
  \Phi_N(\lambda) = N \Phi(\lambda),
\end{equation}
where we take $\Phi(\lambda)$ to indicate the thermodynamic potential of a subsystem.
Note that such a relation can not be written for the entropy.
In the case of entropy one needs to consider the convolution, i.e., also count states where quantities are not evenly distributed over the sub-systems.
Therefore the Laplace transform is a convenient tool here, it transforms convolutions into products.

If one now performs the inverse Laplace transform one finds that
\begin{multline}
  \exp[S_N(X)] = \\ 
  \frac{1}{(2 \pi i)^d} \int \dots \int_{0^+ - i\infty}^{0^+ + i\infty} \exp[ N (-\Phi(\tilde{\lambda}) + \tilde{\lambda} \cdot \bar{X})] d \tilde{\lambda} \\  = \exp[ N (-\Phi(\lambda) + \lambda \cdot \bar{X})] 
\biggl( - N \,\det \Bigl ( 2\pi \, \frac{\partial^2 \Phi(\lambda)}{\partial \lambda \partial \lambda} \Bigr) \biggr)^{- \frac{1}{2}},
\label{eq:inverse-laplace}
\end{multline}
Here $x= X/N$ and the corresponding $\lambda$ follows from eq.~\eqref{eq:equilibrium_X} for $\bar{X}=x$.
This result is obtained by considering the dominant contribution to the integral, i.e. where eq.~\eqref{eq:equilibrium_X} is valid.
There a second order approximation of the integrand is used.
Here we purposely left in the determinant term to illustrate that also here the non-scalar behavior of $S_N(X)$ is apparent.
For large $N$ one finds that
\begin{equation}
  \frac{S_N(X)}{N} = -\Phi(\lambda) + \lambda \cdot x + \mathcal{O}\Bigr(\frac{\ln N}{N}\Bigl ).
\end{equation}
The limit, to a very large system composed of many subsystems,
\begin{equation}
  s(x) = \lim_{N \rightarrow \infty} \frac{S_N(X)}{N}
\end{equation}
is often called ``the thermodynamic entropy''.
We do not want to restrict ourselves to the thermodynamic limit.
Therefore this definition is too restrictive for our purposes.
We will stick to calling $S(X)$ {\em the} entropy.
We will call $s(x)$ the ``thermodynamic-limit entropy''.
The thermodynamic-limit entropy straightforwardly follows from this.

The thermodynamic-limit entropy has the famous properties such as concavity, extensivity etc..
It obeys the ordinary thermodynamic rules.
This thermodynamic potential and $s(x)$ are related by the Legendre transform
\begin{equation}
  s(x) = \lambda \cdot x -\Phi(\lambda),\; x = \frac{\partial \Phi(\lambda)}{\partial \lambda}, \; \lambda = \frac{\partial s(x)}{\partial x}.
  \label{eq:thermodynamics}
\end{equation}
A useful relation, we will use later on, that can be derived from this is
\begin{equation}
  \frac{\partial \lambda}{\partial x} = \left ( \frac{\partial x}{\partial \lambda} \right)^{-1}\; \rightarrow \; \frac{\partial^2 \Phi}{\partial \lambda \partial \lambda} = \left (  \frac{\partial^2 s}{\partial x \partial x} \right)^{-1}.
  \label{eq:second_deriv_s}
\end{equation}

Note that we expressed the entropy, $S$, as function of the extensive quantity $X$.
If we want to express the entropy as function of $x=X/N$, according to the the transformation rule of the entropy, eq.~\eqref{eq:transformed_entropy},
\begin{equation}
  S_N(x) = S_N(X) + d\, \ln N  = N \, s(x) + \mathcal{O} (\ln N).
  \label{eq:thermodynamic_limit_entropy}
\end{equation}
(here $d$ is the dimension of the coarse-grained space).
So the real entropy, also as function of a density, is $N$ times the thermodynamic limit density, also if one uses intensive variables.

The story told seems to be quite generally valid.
It is worthwhile to investigate where it breaks down.
A crucial step takes place at the approximation of the inverse Laplace transform, eq.~\eqref{eq:inverse-laplace}, using the second order approximation of the term in the exponential.
This is not allowed when $\Psi$ is non-analytic at $\lambda$.

When does this occur?
It can occur when $s(x)$ is non-concave, bimodal, for example.
When starting with a non-concave thermodynamic-limit entropy $s(x)$ and inserting $S(X) = N s(X/N)$ into eq.~\eqref{eq:Laplace-transform_S} one finds that
in the limit $N \rightarrow \infty$
\begin{equation}
  \Phi(\lambda) = \inf_x \{\lambda \cdot x - s(x)\}.
\end{equation}
The reason is that the largest term in the exponent dominantly contributes.
This is called the Legendre-Fenchel transform.
When $s(x)$ is concave this transform gives the same result as the Legendre transform.
If $s(x)$ is non-concave only those $x$'s of the domain where $s(x)$ superimposes with the convex-hull of $s(x)$ play a role.
Still, $\Phi(\lambda)$ stays concave, since this is a general property independent of the thermodynamic limit.
However, when computing $\Phi(\lambda)$ using the infimum sometimes the $x$ jumps form one region to the other upon a small change of $\lambda$ to obtain the smallest value.
At these $\lambda$'s there is a non-analyticity in $\Phi(\lambda)$.

Now consider the case that the entropy of a finite subsystem, $S_\mathrm{sub}(X)$, is non-concave.
If $S_\mathrm{sub}(X)$ is well behaved then $Z(\lambda)$ is analytic (and larger than $0$) in $\lambda$.
If one now finds that subsystems are (more of less) independent, then for large $N$, $-N \ln Z(\lambda)$ is also analytic.
As a consequence, $s(x)$, will be found to be concave.
The general conclusion is that entropy $s(x)$ is concave unless the system is non-extensive in the thermodynamic limit.

Put it the other way around, the proof whether a system has an concave thermodynamic limit boils down to proving that the system is extensive.
One can rigorously proof this if, e.g., potentials are sufficiently short ranged \cite{Rue69, Lan73}.
Systems that can show non-concave behavior in the thermodynamic limit are typically systems with long-ranged interactions.
The most notorious class is gravitational systems.
Here one finds non-standard thermodynamics, such as negative heat-capacities \cite{Lyn99}.

For many short ranged systems one has proved that the the entropy is concave.
However, at phase transitions they can behave non-extensively.
In this case systems can be heterogeneous or have long-ranged correlations.
These two conditions can be reconciled by noticing that for these states $s(x)$ has affine patches where
\begin{equation}
  \det \frac{\partial^2 s(x)}{\partial x \partial x} = 0.
\end{equation}
In these cases $s(x)$ is not a good approximations of $S(X)/N$.
When $s(x)$ is very flat, the perturbation due to finite system size, determine the behavior.
It determines whether $S(x)$ behaves bimodal, by means of a ``convex intruder'', or concave.
The non-extensive terms dominate the dynamics and the structure.
For finite size systems in these situations there is difference between the micro-canonical ensemble and the canonical one \cite{Gro05}.
In this case when has to take non-extensive contributions to the entropy into account.
In modeling, this non-extensive behavior is often accounted for by entropy (or free energy) contributions contributed to interfaces or spatial correlations.
Depending on the magnitude of these terms, fluctuations can still be neglected (e.g., two-phase macroscopic flow), or are dominant (critical phenomena).
This is the realm where mesoscopic modeling is often applied.
Because it often hard to model the structure in a macroscopic fashion one remains on a level of description where the structure occurs.

\subsubsection{Interpretation of the entropy definition}

We think it might be helpful to comment on our entropy definition.
Our definition is objective, but depends on the variables $X^i$ used to the describe the system.
It is the logarithm of the density of states (Liouville measure per Lebesgue measure $X$) corresponding to the state $X$.
It counts {\em all} the microscopic states corresponding to a state $X$, so not only states that are sampled in a certain time or something like that.

To illustrate this point let us look at a consequence of this definition.
So let's assume that the coarse-grained state $X$ contains information on number of particles, momentum etc.
In this case the number of particles in a certain volume $V_i$ can be computed from $X$ to be $M_i(X)$ ($i=1\cdots N$).
Now consider the macroscopic entropy definition.
Let $\Gamma_\mu$ be the phase space coordinates of particle $\Gamma_\mu$ (particle position and momentum).
Assume that the macro state does not change under the interchange of particles, so $X(\dots, \Gamma_\mu ,\dots, \Gamma_\nu,\dots)= X(\dots, \Gamma_\nu ,\dots, \Gamma_\mu,\dots)$.
When computing $\exp[S(X)]$ one needs to integrate $\Gamma_\mu$ over all the spatial and momentum domain.
Using the symmetry of $X$ this can be simplified,
\begin{multline}
    \exp[S(X)] = \int_{\cup_i V_i} \prod_{\mu=1}^M d\mu_L(\Gamma_\mu) \, \delta[ X -X(\Gamma)] \\
    = \left( \prod_i \int_{V_i} \prod_{\mu=1+CM_{i-1}}^{yM_i} d\mu_L(\Gamma_\mu) \right) \, \frac{M! \delta[ X -X(\Gamma)]}{M_1(X)! \cdots M_N(X)!},
\end{multline}
i.e., one restricts the integration by $M_i$ particles per volume ($CM_i$ denotes the cumulative sum up to $i$) and accounts for all possible permutations by means of the multinomial coefficient.
Note that if the subsystems are sufficiently independent, such that the thermodynamic limit is valid for each cell individually, then
\begin{equation}
  \frac{1}{M!} \exp[S(X)] \approx \prod_i \frac{1}{M_i(X)!} \exp[S(X^i)].
  \label{eq:extensive_local_entropy}
\end{equation}
The full state of the system $X$ is characterized by the states of the subsystems $X=(X^1, \dots, X^i, \dots)$.
In this case $X^i$ can be the extensive state of cell $i$, e.g., the total particle number, momentum and energy associated with cell $i$.
Here we use ``associated'' because not all quantities are fully localized.
For example, if one has pair-interactions, part of the energy is due to interaction of particles in neighboring cells.
Now one could account for this interaction energy by associating half of it with each of the neighboring cells.

This explains the $1/M_i!$ factor in the usual entropy definition.
The treatment as presented here resolves the Gibbs paradox.
When considering more then one species of particles, one introduces a distinction between these variables by means of $X$.
One can, for example, use the density of the species.
This means one that $X$ is only invariant under exchange of particles of the same species.
Therefore the combinatorial factor is different.
For a discussion of the Gibbs paradox and a solution of it along similar lines see \cite{Jay92}.

It is probably also clarifying to command on the status of the Gibbs entropy within this framework.
This entropy, under the name of relative entropy, arises naturally in the context of large-deviation theory.
It is only valid in the thermodynamic limit.
This point of view is easiest explained in the discrete case.
Let's consider a larger number, $N$, of independent subsystems and discrete states.
Now $\rho_\alpha$ counts the fraction of subsystems with state $\alpha$.
The total entropy of a state $X=N \sum_{\alpha=1}^N \rho_\alpha \, X_\alpha$ (here $X_\alpha$ is the value $X$ corresponding to a state $\alpha$ and not the state of a subsystem with index $\alpha$).
By counting the number of subsystems in the same state one can express the entropy by considering all permutations of subsystems in different states,
\begin{multline}
  \exp[S(X)] = \sum \frac{N!}{ \prod_\alpha (N \rho_\alpha)! }  \, \mathbf{1}_0(X - N \sum_\alpha \rho_\alpha X_\alpha) \\\times \exp[ N \sum_\alpha \rho_\alpha S(X_\alpha)].
\end{multline}
The Gibbs entropy is an approximation of the multinomial factor.
Consider the Gibbs entropy,
\begin{multline}
  S_G = - \sum_\alpha \rho_\alpha \ln (\rho_\alpha \exp[-S(X_\alpha)])\\
  \approx \frac{1}{N} \ln \left( \frac{N!}{ \prod_\alpha (N \rho_\alpha)! } \right) + \sum_\alpha \rho_\alpha S(X_\alpha).
\end{multline}
(in the discrete case $\exp[S_\alpha]$ corresponds to the Liouville measure corresponding to the discrete state $\alpha$).
Maximizing $S_G$ under the constrained that $X = N \sum_\alpha \rho_\alpha X_\alpha$ gives the dominating term in the sum.

The computation makes no sense of the ensemble is purely fictional.
There have to be real possibilities to distribute the extensive quantity over $N$ subsystems.
A similar interpretation of the different entropy definitions can be found in \cite{Pen69} and \cite{Leb99}.

\subsection{Expectation values in the thermodynamic limit}

Let us consider the case where,
\begin{equation}
  \begin{split}
  x(\Gamma) &= \frac{1}{N} \sum_\alpha x_\mathrm{sub}(\Gamma^\alpha)\\
  a(\Gamma) &= \frac{1}{N} \sum_\alpha a_\mathrm{sub}(\Gamma^\alpha).
  \end{split}
\end{equation}
For large $N$ we want to know the expectation value $\mathbb{E}(a|x)$.
To compute this value we can use the Laplace-transform trick:
\begin{multline}
\int \exp[S(x)] \, \mathbb{E}(a|x) \, \exp[ -N \, \lambda \cdot x] \, dx \\
= Z_\mathrm{sub}(\lambda)^N \,Z_\mathrm{sub}(\lambda)^{-1} \int a_\mathrm{sub}(\Gamma^\alpha) \, \exp[ - \lambda \cdot x_\mathrm{sub}(\Gamma^\alpha)] \, \mu_L[\Gamma^\alpha]\\
= \exp[- N \, \Phi(\lambda)] \; \langle a_\mathrm{sub} ,  \mu^c(\lambda) \rangle
\end{multline}
Here eq.~\eqref{eq:conditional_expectation} was used.
The measure $\mu^c(\lambda)$ indicates the generalized canonical probability measure
\begin{equation}
  d\mu^c(\lambda)[\Gamma^\alpha] = Z_\mathrm{sub}^{-1}(\lambda) \, \exp[ - \lambda \cdot x_\mathrm{sub}(\Gamma^\alpha)] \, d\mu_L[\Gamma^\alpha].
\end{equation}

Inverting this relation for large $N$ and analytic $\Phi(\lambda)$, gives using the inverse Laplace-transform a concentration for $\lambda$ that is related to $x$ by means of eq.~\eqref{eq:thermodynamics}.
For $N\rightarrow \infty$ the expectation value is therefore well approximated by the canonical expectation-value,
\begin{equation}
  \mathbb{E}(a|x) = \langle a_\mathrm{sub} ,  \mu^c(\lambda) \rangle \equiv \langle a_\mathrm{sub} \rangle_\lambda
\end{equation}
The deviation is $\mathcal{O}(N^{-1})$.
Below, we will also need to consider the situation of expectation value of a product, $\mathbb{E}(a \, b|x)$, where $b$ is of a similar form as $a$.
In this case one finds that
\begin{multline}
  N^{-2} \sum_\alpha \sum_\beta \int a_\mathrm{sub}(\Gamma^\alpha) \, b_\mathrm{sub}(\Gamma^\alpha)  \\ \times \exp[ - \lambda \cdot (x_\mathrm{sub}(\Gamma^\alpha)+x_\mathrm{sub}(\Gamma^\beta)) ] \, \prod_\gamma d\mu_L[\Gamma^\gamma]\\
 = Z_\mathrm{sub}(\lambda)^N \Bigl( N^{-1} \langle a_\mathrm{sub} \, b_\mathrm{sub} \rangle_\lambda   \\ + (1-N^{-1}) \,  \langle a_\mathrm{sub}\rangle_\lambda \, \langle b_\mathrm{sub}\rangle_\lambda\Bigr)
\end{multline}
As a result we find that
\begin{equation}
  \begin{split}
    \mathbb{E}(a \, b|x)  &= \langle a_\mathrm{sub} \rangle_\lambda \, \langle b_\mathrm{sub} \rangle_\lambda + \mathcal{O}(N^{-1})\\
    &= \mathbb{E}(a |x)\, \mathbb{E}(b|x) + \mathcal{O}(N^{-1}).
  \end{split}  
  \label{eq:product-expectation}
\end{equation}

\subsection{Successive Coarse-graining of extensive systems}

We will consider systems that are well inside the thermodynamic limit.
In general, assuming that the entropy (near the thermodynamic limit) obeys relation eq.~\eqref{eq:extensive_local_entropy} is not a good idea.
In many situations, coarse-grained variables are not neatly localized in non-overlapping cells.
One might think of the situation where one represents a state using base functions, such as in the finite element method, or in a Fourier representation.
What we will assume is that on some, relatively, fine scale a system can be written by a suitable choice of variables in the form of eq.~\eqref{eq:extensive_local_entropy}.

The level we are considering is a coarse-graining of this (thermodynamic limit) finer level.
At such a coarse-grained level, one can approximate the entropy well by
\begin{equation}
  S(x) = N \, s(x),
  \label{eq:thermodynamic-limit_entropy}
\end{equation}
where $N$ can be considered a large variable.
Differently from \S\ref{sect:therm_limit} where a homogeneous situation was considered here elements $x^i$ can denote the state of well separated regions in space, or states corresponding to different wave-vectors etc.
The reason this expression is valid, is because it can be derived from coarse-graining a a finer scale that obeys eq.~\eqref{eq:extensive_local_entropy}.
On the finer level the equation is valid for the individual subsystems (that themselves can be divided in $N$ independent subsystems).
Below we will proof that once the entropy has the form eq.~\eqref{eq:thermodynamic-limit_entropy} it will remain of this form upon coarse-graining.

In the thermodynamic limit degeneracy conditions simplify, for example eq.~\eqref{eq:isentropic}, becomes
\begin{multline}
  \exp[N \, s(x)] \, \frac{\partial}{\partial x} \cdot \Bigl ( \exp[N \, s(x)] \, \dot{x}^\mathrm{rev} \Bigr) = \\ \frac{\partial}{\partial x} \dot{x}^\mathrm{rev} + N \, \dot{x}^\mathrm{rev} \cdot \frac{\partial}{\partial x} =0 \; \rightarrow \; \dot{x}^\mathrm{rev} \cdot \frac{\partial s}{\partial x} =0,
\end{multline}
because $N$ is extremely large.
Therefore the (generalized) Liouville theorem gives rise to the isentropic condition for reversible motion in the thermodynamic limit.
 
When successively coarse-graining this state further we will consider states $y=y(x)$.
It is convenient to work with intensive variables, such as mass, momentum and energy densities.
The reason is that  when the variation is little, the coarse-grained values are close to the original ones.
The entropy of this coarse-grained scale then follows as,
\begin{equation}
  \begin{split}
    \exp[S(y)] &= \int \delta[y-y(x)] \exp[S(x)] \, dx\\
    & \approx \int \delta[y-y(x)] \exp[N \, s(x)] \, dx\\
    & \approx \int \delta\biggl[ (x-\bar{x})\cdot \frac{\partial y(\bar{x})}{\partial \bar{x}} \biggr] \, \exp\biggl[N \, s(\bar{x}) \\ & \quad + N \frac{\partial^2 s(\bar{x})}{\partial \bar{x} \partial \bar{x}} : (x-\bar{x}) (x-\bar{x})\biggr ] \, dx\\
    S(y) &= N \, s(\bar{x}) + \mathcal{O}(\ln N) \; \rightarrow \; s(y) = s(\bar{x}).
  \end{split}
\end{equation}
This proofs that the entropy remains of the shape eq.~\eqref{eq:thermodynamic-limit_entropy}.
The integral is dominated by the maximum thermodynamic-limit entropy under the constraint that $y(x)=y$.
Because $s(x)$ is concave this maximum is reached at a unique value of $\bar{x}=x$.
The value can be found by determining the unique values of $\bar{x}$ and $\lambda^y$, for which two equations hold simultaneously, namely
\begin{equation}
  \frac{\partial s(\bar{x})}{\partial \bar{x}} = \lambda^y(y) \cdot \frac{\partial y(\bar{x})}{\partial \bar{x}}, \text{ and } y(\bar{x}) = y.
\end{equation}
Using the Legendre transform for the entropy to the thermodynamic potential, $\Phi(x)= \lambda \cdot x - s(x)$, the equations can also be restated as
\begin{equation}
  \bar{\lambda} = \lambda^y \cdot \frac{\partial y(\bar{x})}{\partial \bar{x}}\text{ and } y(\lambda^y)=y.
\end{equation}
For this state $\bar{x}$ corresponding to the maximum (constrained) entropy we thus have $s(y)=s(\bar{x})$.
The relation between coarse-grained thermodynamic potentials is a bit less straightforward,
\begin{equation}
  \Phi^y(\lambda^y) = \lambda^y \cdot y - s(y)= \Phi(\bar{\lambda}) + \lambda^y \cdot \Bigl( y - \frac{\partial y(\bar{x})}{\partial \bar{x}} \cdot \bar{x} \Bigr).
  \label{eq:coarse-grained-pot}
\end{equation}
An more elegant relation, that will be used below, between the coarse-grained thermodynamic potentials follows from the definitions as,
\begin{equation}
  \frac{\partial^2 \Phi^y}{\partial \lambda^y \partial \lambda^y} = \frac{\partial y}{\partial \bar{x}} \cdot \frac{\partial^2 \Phi}{\partial \bar{\lambda} \partial \bar{\lambda}} \cdot \frac{\partial \bar{\lambda}}{\partial \lambda^y}.
\end{equation}
Using eq.~\eqref{eq:second_deriv_s} this relation can also be used to relate the (inverse) second derivatives of the thermodynamic-limit entropy.
The expression for the derivative of the thermodynamic driving forces $\lambda$ is a bit more troublesome, namely,
\begin{equation}
  \frac{\partial \bar{\lambda}}{\partial \lambda^y} = \Bigl ( 1-\lambda_c \cdot \frac{\partial^2 y}{\partial \bar{x} \partial \bar{x}} \cdot \frac{\partial^2 \Phi}{\partial \bar{\lambda} \partial \bar{\lambda}} \Bigr)^{-1} \cdot \frac{\partial y}{\partial \bar{x}},
\end{equation}
resulting, using eq.~\eqref{eq:second_deriv_s}, in
\begin{equation}
  \begin{split}
    \frac{\partial^2 \Phi^y}{\partial \lambda^y \partial \lambda^y} &= \frac{\partial y}{\partial \bar{x}} \cdot \Bigl (\frac{\partial^2 s}{\partial \bar{x} \partial \bar{x}} - \lambda_c \cdot \frac{\partial^2 y}{\partial \bar{x} \partial \bar{x}} \Bigl )^{-1} \cdot \frac{\partial y}{\partial \bar{x}}\\
    &=\Bigl (\frac{\partial^2 s}{\partial y \partial y}\Bigr)^{-1}.    
  \end{split}
  \label{eq:coarse-grained-pot_2nd_deriv}
\end{equation}

Both expressions, eq.~\eqref{eq:coarse-grained-pot} and eq.~\eqref{eq:coarse-grained-pot_2nd_deriv} are complicated by the occurrence of the second derivative of $y(x)$.
In practice most transformations, in the thermodynamic limit, are linear.
The reason is that one looks at coarse-graining extensive variables (or densities).
In this case only weighted averaging, which is a linear transformation, as coarse-graining makes really sense.
Therefore we will assume, in the following, that
\begin{equation}
  \frac{\partial y}{\partial x} = \ten{B},
\end{equation}
is a matrix independent of $x$.
If one should an encounter a situation where this assumption is not valid the analysis below should be redone including the complication.
This is straightforward, but a little bit more involved.

In the thermodynamic limit expectation values are concentrated.
Expectation values are dominated by the maximum thermodynamic-limit entropy, obeying the constraint $y=y(\bar{x})$, so for functions $A(x)$,
\begin{equation}
  \mathbb{E}(A|y) = \mathbb{E}(A|\bar{x}) = A(\bar{x}).
\end{equation}
For the special case of the instantaneous rate of change, one finds that
\begin{equation}
  \mathbb{E}(\dot{y}|y) = \ten{B} \cdot \mathbb{E}(\dot{x}|\bar{x}).
\end{equation}
If one wants to express the result using $\ten{\Omega}$, eq.~\eqref{eq:coarse-graining_Omega}, one gets
\begin{equation}
  \ten{\Omega}^y(y) = \ten{B} \cdot \ten{\Omega}(\bar{x}) \cdot \ten{B}.
\end{equation}

If the energy is extensive (if interactions are short-ranged), one can use eq.~\eqref{eq:product-expectation}, on the definition of $\ten{\Omega}$, eq.~\eqref{eq:Omega_def}.
One finds that
\begin{equation}
  \ten{\Omega}(x) = \ten{L}(x) \cdot \mathbb{E}(H|x) + \mathcal{O}(N^{-1}).
\end{equation}
Let us name this $\mathcal{O}(N^{-1})$-term:
\begin{equation}
  \tilde{\ten{\Omega}}(x) \equiv  E \biggl(\frac{\partial x}{\partial \Gamma} \cdot \ten{L}_\mathrm{micro}  \cdot \frac{\partial x}{\partial \Gamma} \, \bigl (H(\Gamma)-\mathbb{E}(H|x) \bigr ) \biggl |x \biggr).
  \label{eq:Omega_tilde}
\end{equation}
This term is zero if the total energy can be expressed in terms of the coarse-grained variables, i.e., if $H=\mathbb{E}(H|x)$.
Using this decomposition we obtain, from eq.~\eqref{eq:instantaneous}, that
\begin{equation}
  \mathbb{E}(\dot{x}|x) = \ten{L} \cdot \frac{\partial}{\partial x} \mathbb{E}(H|x) + \exp[-S(x)] \frac{\partial}{\partial x} \cdot \Bigl( \tilde{\ten{\Omega}}^T \exp[S(x)] \Bigr ).
\end{equation}
Here we used property eq.~\eqref{eq:degeneracy_L}.

If a cell consists of $N$ independent subsystems the energy is extensive (if interactions are short ranged).
The microscopic $\ten{L}$ matrix can be well approximated by means of a block-diagonal form.
Since we are interested in scaling behavior we will assume that subsystems are fully statistically independent,
\begin{equation}
  \begin{split}
    \Gamma &= (\Gamma^1, \dots, \Gamma^N),\\ 
    x(\Gamma)&= \frac{1}{N} \sum_{\alpha=1}^N x(\Gamma^\alpha),\\
    H(\Gamma) &= \sum_{\alpha} H(\Gamma^\alpha) = N^{-1} \sum_{\alpha} N \, H(\Gamma^\alpha)\\
    \ten{L}_\mathrm{micro}(\Gamma) &= \ten{L}_\mathrm{micro}(\Gamma^1) \oplus \cdots \oplus \ten{L}_\mathrm{micro}(\Gamma^N)
  \end{split}
\end{equation}
This gives that
\begin{equation}
  \begin{split}
  H(x) = N \, h(x) \text{ and } \ten{L}(x) = N^{-1} \, \ten{l}(x).
  \end{split}
\end{equation} 
Combining this with the earlier observations that
\begin{equation}
  S(x) = N \, s(x) \text{ and } \tilde{\ten{\Omega}}(x) = N^{-1}  \tilde{\ten{\omega}}(x),
\end{equation}
we find that
\begin{equation}
  \mathbb{E}(\dot{x}|x) = \ten{l} \cdot \frac{\partial h}{\partial x} + \tilde{\ten{\omega}} \cdot \frac{\partial s}{\partial x} + \mathcal{O}(N^{-1}).
\end{equation}

In this thermodynamic limit the conditions, eq.~\eqref{eq:degeneracy} and eq.~\eqref{eq:degeneracy_L}, reduce to
\begin{equation}
  \mathbb{E}(\dot{x}|x)\cdot \frac{\partial s}{\partial x}=0 \text{ and }  \ten{l} \cdot \frac{\partial s}{\partial x} =0.
\end{equation}
The instantaneous rate of change is isentropic in the thermodynamic limit.
Because of the anti-symmetry of $\tilde{\ten{\omega}}$ this term also is isentropic, and the second equation implies the first one.

We expect that, in eq.~\eqref{eq:dissipation}, only local contributions correlate.
In the thermodynamic limit we therefore expect that $\tilde{\ten{M}} = \tilde{\ten{m}}/N$ is the appropriate scaling with $N$ in the thermodynamic-limit regime.
Using this relation we obtain that
\begin{equation}
    \exp[-N\, s(x)] \frac{\partial}{\partial x} \cdot \Bigl (\exp[N\, s(x)] \, \tilde{\ten{M}}^T \Bigr) \approx \tilde{\ten{m}} \cdot \frac{\partial s}{\partial x}.
  \label{eq:thermodynamic-limit_M}
\end{equation}
Where the equation is exacts in the thermodynamic limit.
When we insert this relation into eq.~\eqref{eq:coarse-grained-fluctuations} with $A=x$ this gives

\begin{widetext}
\begin{multline}
    \Delta x^{y,\mathrm{fluct}}_t =\int_0^t \Bigl ( \ten{l}(x^{y,\mathrm{fluct}}_{t'}) \cdot \frac{\partial h}{\partial x^{y,\mathrm{fluct}}_{t'}} - \ten{l}(\bar{x}^{y,\mathrm{fluct}}_{t'}) \cdot \frac{\partial h}{\partial \bar{x}^{y,\mathrm{fluct}}_{t'}}\Bigr ) d{t'}
+  \int_0^t \Bigl ( \tilde{\ten{\omega}}(x^{y,\mathrm{fluct}}_{t'}) \cdot \frac{\partial s}{\partial x^{y,\mathrm{fluct}}_{t'}} - \tilde{\ten{\omega}}(\bar{x}^{y,\mathrm{fluct}}_{t'}) \cdot \frac{\partial s}{\partial \bar{x}^{y,\mathrm{fluct}}_{t'}}\Bigr ) d{t'}\\
     + \int_0^t \Bigl ( \tilde{\ten{m}}_{t-t'}(x^{y,\mathrm{fluct}}_{t'}) \cdot \frac{\partial s}{\partial x^{y,\mathrm{fluct}}_{t'}} 
      - \tilde{\ten{m}}_{t-t'}(\bar{x}^{y,\mathrm{fluct}}_{t'}) \cdot \frac{\partial s}{\partial \bar{x}^{y,\mathrm{fluct}}_{t'}} \Bigr ) \, d{t'} + \Delta x^\mathrm{fluct}_t.
\end{multline}
\end{widetext}

Note that $\bar{x}$ can always be computed from $x$, since $\bar{x}$ follows from $y(x)$.
Therefore the equation is a closed equation.
This equation tells us that $x^{y,\mathrm{fluct}}_t$ will be driven toward $\bar{x}^{y,\mathrm{fluct}}_t$.
The difference between the two is kept away from zero by $\Delta x^\mathrm{fluct}_t$.
Since this fluctuating contribution approaches $0$ for $N\rightarrow \infty$ it can be considered a perturbation.
We expect $\Delta x^\mathrm{fluct}_t = \mathcal{O}(N^{-\frac{1}{2}})$, which is dominant compared to other $\mathcal{O}(N^{-1} \ln N)$ contributions.
Therefore $\Delta x^\mathrm{fluct}_t$ is the only perturbation that needs to be considered.
We can make a first order Taylor expansion around the initial value $\bar{x}^{y,\mathrm{fluct}}_0=\bar{x}_0$, i.e., $x^{y,\mathrm{fluct}}_t= \bar{x}_0+ \delta x_t$,
\begin{multline}
  \Delta \delta x_t  = \frac{\partial }{\partial \bar{x}_0} \Bigl ( \ten{l}(\bar{x}_0) \cdot \frac{\partial h}{\partial \bar{x}_0}\Bigr) \cdot \int_0^t  (\delta x_{t'} - \delta \bar{x}_{t'}) \, d{t'}\\
 + \frac{\partial }{\partial \bar{x}_0} \Bigl ( \tilde{\ten{\omega}}(\bar{x}_0) \cdot \frac{\partial s}{\partial \bar{x}_0}\Bigr) \cdot \int_0^t  (\delta x_{t'} - \delta \bar{x}_{t'}) \, d{t'}\\
     \quad + \int_0^t \frac{\partial }{\partial \bar{x}_0} \Bigl ( \tilde{\ten{m}}_{t-t'}(\bar{x}_0) \cdot \frac{\partial s}{\partial \bar{x}_0} \Bigr ) \cdot (\delta x_{t'} - \delta \bar{x}_{t'}) \, d{t'} \\ + \Delta x^\mathrm{fluct}_t.
\end{multline}
The value of $\delta \bar{x}_{t}$ follows from $\delta x_{t}$ as,
\begin{equation}
  \delta \bar{x}_{t} = \ten{\Lambda}^{-1} \cdot \ten{B} \cdot \ten{\Lambda}^y  \cdot \ten{B} \cdot \delta x_{t},
\end{equation}
where the definitions,
\begin{equation}
  \ten{\Lambda}(\bar{x}_0) = \frac{\partial^2 s}{\partial \bar{x} \partial \bar{x}} \text{ and } \ten{\Lambda}^y(\bar{x}_0)= \frac{\partial^2 s}{\partial y \partial y} = \Bigl( \ten{B} \cdot \ten{\Lambda}^{-1} \cdot \ten{B} \Bigr )^{-1},
\end{equation}
are as in the linear case \S\ref{sec:linear}, except that here the matrices depend on the parameter $\bar{x}_0$.
This parameter is a constant in the equation for $\delta x$.
Therefore the solution is completely equivalent as for the linear case.
We obtain for $\ten{m}^{y,x}_{t}(\bar{x}_0) = N \, E (\delta x_t \delta \dot{x}_0|\bar{x}_0)$, in the Laplace-transformed form,
\begin{multline}
  \ten{m}^{y,x}_s(\bar{x}_0) = 
\Biggl [1 + \biggr ( (s^{-1} \tilde{\ten{\omega}} + \tilde{\ten{m}}_s) + 
 \Bigl( \bar{\lambda}_0 \cdot \frac{\partial (s^{-1} \, \tilde{\ten{\omega}} + \tilde{\ten{m}}_s)}{\partial \bar{x}_0} \\+ \frac{\partial }{\partial \bar{x}_0} \Bigl (s^{-1} \, \ten{l}(\bar{x}_0) \cdot \frac{\partial h}{\partial \bar{x}_0}\Bigr) \Bigr ) \cdot \ten{\Lambda}^{-1} \biggr) \cdot \ten{\tilde{\Lambda}} \Biggr ]^{-1} \cdot \tilde{\ten{m}}_s(\bar{x}_0),
\label{eq:Laplace-nonlinear}
\end{multline}
For times large enough, such that the short time-scales corresponding to the poles of eq.~\eqref{eq:Laplace-nonlinear} are decayed to zero,
one thus finds the approximation, similar to eq.~\eqref{eq:M_c-linear} (or better eq.~\eqref{eq:M_c-linear2})
\begin{equation}
  \ten{m}^{y}(y)= (\ten{\Lambda}^y)^{-1}(y) \cdot \ten{G}^{-1}(\bar{x}) \cdot (\ten{\Lambda}^y)^{-1}(y). 
\end{equation}
These results are fully rigorous in the thermodynamic limit (except for the restriction that $\ten{B}$ is taken constant).

\section{Conclusions and Discussion}
\label{sect:concl}

We performed a careful derivation of the generalized Langevin equation.
Performing this derivation we found that the Zwanzig formalism is superior compared to others (Mori, Robertson, Grabert).
The reason is that the canonical ensemble is not well suited in the case where fluctuations are dominant.
To be able to perform non-equilibrium predictions using the Robertson and Grabert formalism one needs to perform quite artificial adaptations of the Mori formalism.
The Zwanzig flavor, projection operator formalism, does not have these problems.
The important underlying reason is that this projection is optimal in a sense of optimal prediction theory.

The derivation, as a result, gives a microcanonical entropy definition.
This definition is objective and depends on which macroscopic variables, $X$, are used.
The entropy is the logarithm of the density of states (Liouville measure) per unit volume $X$ (Lebesgue measure).
To compute the entropy, one should take into account all microstates consistent with a macro state and not only the states actually sampled.
Entropy arises in dynamic equations because it measures the amount of phase space available when a system changes its coarse-grained state $X$.
If there is more phase-space available there is a bias to go to that state.
This is the thermodynamic driving force.
The ergodic point of view that entropy has to do with phase space visited in a certain time is not supported by our analysis.

To illustrate this statement let us consider the entropy of a high molecular weight, entangled, polymer melt.
Upon deformation the polymer chains gets stretched (on average).
Subsequently the polymer conformations will try to relax towards equilibrium.
Initially this relaxation is quick but soon polymer molecules will start feeling their neighbors.
Because the melt is entangled relaxations slows down.
According to the theory of Doi and Edwards \cite{Doi86} conformations will be confined to a tube-like region.
The contour-length and the cross sectional area of the tube is independent of the deformation.
A polymer can only relax further by escaping the tube (so-called reptation).
So, there is a two step process of relaxation, namely, a fast process of the chain inside the tube and a slow one of the tube itself.
here is a big gap between the characteristic time scales.

Here comes the point.
Suppose after a step-strain and subsequent fast relaxation inside the tube one characterizes the state by the strain.
One want to know the entropy as a function of the strain.
One might think the entropy can be computed from the number chain conformations sampled by a chain inside each tube.
Since the contour length and radius of the tube is not change after deformation (and subsequent relaxation) one finds this phase space volume is independent of strain.
The mistake is that, in fact, also the number of tube configurations, consistent with the strain-deformation should be taken into account, although these conformation are almost static on the time-scale under consideration.
All entropy comes from this contribution.

Entropy is not a scalar quantity.
So, upon a change of variables extra terms appear.
In the thermodynamic limit these terms are negligible.
It has, however, consequences for small systems.
In this case the current entropy definition deviates from other ones such as the Gibbs entropy.
Because of the rigorous connection through Zwanzig projection operator formalism with microscopic dynamics the current entropy definition is proved to be the correct one to use.
If one approximates the governing equation with a stochastic differential equation the non-scalar transformation rule is essential.
Only when allowing the entropy to transform in this way the form of the equation does not change upon a change of coordinates, as follows from Ito-calculus.
The Langevin equation poses no restriction on the set of variables one uses to describe a system.
The choice should be motivated by the problem at hand.
What determines a good choice is the decorrelation behavior of fluctuations of the macroscopic variables.
If they decorrelate quickly the formal generalized Langevin equation can be approximated by a practically useful stochastic equations.
Only the generalized Langevin equation can be rigorously approximated by a stochastic differential equation.
The reason is that the fluctuating contributions can be seen as a path through the $X$-space, such that $A^\mathrm{fluct}_t=A(X^\mathrm{fluct}_t)$ for functions $A(X)$.
This equality does not hold for the other projection-operator flavors.

We provided the equations of successive coarse-graining.
A stochastic differential equation for the coarse-grained variable $Y$ can be provided if one knows the statistics of its fluctuations $Y^{y,\mathrm{fluct}}$.
Besides giving the general equation governing these fluctuations, eq.~\eqref{eq:coarse-grained-fluctuations}, we studied and solved it for the linear regime and the thermodynamic-limit case.
The general picture that follows from this is the following.
The only fluctuations in the fine-scale, $X$, that are important to determine $\ten{M}^y$ are fluctuations that do change $Y$.
Fluctuations that do not influence $Y$ will stay close to $\bar{X}$.
Here $\bar{X}$ is the maximum entropy, $S[X]$, that obeys the constraint $Y(X)=Y(\bar{X})$.
These irrelevant fluctuations are filtered out by a matrix $\ten{Q}$.
We gave an exact recipe how to compute this matrix.

The procedure also indicates for which time-scales the coarse-grained equation is expected to be valid.
Note that when motion of $X$ out of the $Y(X)$ is only slowly relaxing these times can be very long.
Introducing new variables that catch this slow mode are then beneficiary.
This is the art of coarse-graining.

In several treatments, e.g. \cite{Esp02, Ott05}, expressions for successive coarse-graining are presented for stochastic differential equations.
Typically, somewhere in these derivation $X^{y,\mathrm{fluct}}_t$ is replaced by $X^{\mathrm{fluct}}_t$.
The argument used for making this assumption is that $Y$ is a very slow variable.
When obtaining $\ten{M}^y$ in this way one should only consider correlations on time-scales where $Y$ is indeed slow.
As we showed in our derivation, however, to find the full $\ten{M}^y$ one needs to wait until all irrelevant fluctuations have relaxed.
Therefore, when there is no wide separation of time-scale, methods like those provided in \cite{Esp02, Ott05} will not give accurate results.

In our procedure we give an exact equation for $X^{y,\mathrm{fluct}}_t$.
In this equation the thermodynamic driving force that relaxes $Y$ is eliminated.
Therefore long $t\rightarrow \infty$ can be evaluated. 
Our relations reduce in a certain limit to homogenization theory, but they are more general.
When transport-coefficients are derived in the thermodynamic limit they exhibit, usually, long time-tales.
Our method provides a cutoff that depends on the choice of coarse-grained variables $Y$.
We believe that our method provides a cutoff-time for this tail.
Below the cutoff the tail will be present and be part of the transport coefficient.
Above the cutoff time the physical phenomena that causes the tail is part of the coarse-grained equation.

The motivation of this research came from the need for coarse-graining method in computational methods.
When coarse-graining form a molecular to a mesoscopic level, usually, thermodynamic-limit assumptions are not valid.
We therefore hope that the method outlined in this paper will help to bridge the scales in these kinds of methods.

A second application we have in mind is coarse-graining as an alternative for discretization.
The coarse-graining procedure gives a recipe to generate equations using a finite number of variables.
This is typically what is done when performing a discretization.
Because, however, coarse-graining is physical we do expect it may give rise to more stable methods.
The thermodynamics is obeyed, so reversible parts are isentropic and irreversible part entropy increasing (in the absence of fluctuations).
An obvious framework to apply the method to is stabilization in the finite element method \cite{Hug98}. 

\appendix

\section{Mori and Robertson/Grabert projection operators}
\label{app:proj-op}

The goal is to device projection operators that link the microscopic description of a system to a coarser description.
Instead of characterizing a system with a microscopic state $\Gamma$ one would like to use a macroscopic (or mesoscopic) state $X$.
The space of macro states is assumed to be much lower dimensional.
A macro state characterizes a subspace of the microscopic space namely sets of microstates where $X(\Gamma)$ have the same value.

To make a direct link with statistical mechanics, in the Mori and Robertson/Grabert formalism, one tries to express projections to the macro state as an expectation value over an ensemble.
To link microscopic and macroscopic spaces a relevant (probability) measures, $\mu^\mathrm{rel}$ is introduced.

The microstates are assigned a portion of the statistical weight of the micro state. Typically (but not necessarily) generalized canonical ensembles are used.
In this case
\begin{equation}
  d\mu^\mathrm{rel}(X)[\Gamma] = \frac{\exp[ - \lambda(X) \cdot X(\Gamma)]}{Z(X)} \, d\mu_L[\Gamma]. 
  \label{eq:canonical}
\end{equation}
The convention we will use is the following.
When using this measure in an integral the value in the round brackets is fixed while $\Gamma$ is integrated over.
So, when integrating over the microscopic space, $\Gamma$ the $\lambda(X)$ and $Z(X)$ are fixed, but $X(\Gamma)$ varies with $\Gamma$.
Following the usual frame work for canonical ensembles one has $Z(X) \equiv Z(\lambda(X))$, where
\begin{equation}
Z(\lambda) = \int \exp[ - \lambda \cdot X(\Gamma)] \, d\mu_L[\Gamma].
\end{equation}
The functional relations of $\lambda$ on $X$ are such that
\begin{equation}
  X = \langle X, \mu^\mathrm{rel}(X) \rangle = \int X(\Gamma) \, d\mu^\mathrm{rel}(X)[\Gamma],
\end{equation}
which gives the usual
\begin{equation}
  X(\lambda) = - \frac{\partial \ln Z(\lambda)}{\partial \lambda}, \text{ and } \lambda(X) = \frac{\partial \ln S^\mathrm{C}(X)}{\partial X},
\end{equation}
where
\begin{equation}
  S^\mathrm{C}(X) = \lambda(X)\, X + \ln Z(X).
\end{equation}

One might be tempted to interpret the expectation value of $A(\Gamma)$ with respect to $\mu^\mathrm{rel}(X)$ as a projection of $A$ onto $X$, 
\begin{equation}
  (\mathcal{P}^\mathrm{C} A) (X) \equiv \langle A, \mu^\mathrm{rel}(X) \rangle = \int A(\Gamma) \, d\mu^\mathrm{rel}(X)[\Gamma].
\end{equation}
However, $\mathcal{P}^\mathrm{C}$ is not a projection-operator, since
\begin{multline}
(\mathcal{P}^\mathrm{C} \mathcal{P}^\mathrm{C} A)(X) = \langle (\mathcal{P}^\mathrm{C} A )(X), \mu^\mathrm{rel}(X) \rangle = \\
    \int (\mathcal{P}^\mathrm{C} \, A)(X(\Gamma)) \, d\mu^\mathrm{rel}(X)[\Gamma] \ne (\mathcal{P}^\mathrm{C} \, A)(X),
\end{multline}
in general!
Therefore the canonical expectation value can not be interpreted as a projection (which, by definition, obeys $\mathcal{P}^2 = \mathcal{P}$).

One can make either of two choices if one wants to proceed.
The first is to keep $\mathcal{P}^\mathrm{C}$ and accept it is not a projection operator.
Using the formalism and decomposing the equation according to eq.~\eqref{eq:generalized_langevin} gives a fluctuating term that does not obey eq.~\eqref{eq:projected_fluctuations}.
Clearly, this is not done within the projection operator formalism.
Eq.~\eqref{eq:projected_fluctuations} is the main identity that is used to make subsequent approximations.
Therefore this choice is not to be preferred.
A second choice is to change $\mathcal{P}^\mathrm{C}$ a bit such that the operator becomes a projection-operator.
In this case the projection can only be identified approximately as an expectation value over the canonical ensemble.
Therefore part of the link to equilibrium statistical mechanics is lost. 
The second choice is, in our opinion, a less than elegant fix.

One fix for $\mathcal{P}^\mathrm{C}$ to is to linearize $\mu^\mathrm{rel}(X)$ with respect to $X$.
This is done both in the Mori and the Robertson flavor of the projection operator formalism.
In the Mori flavor one linearizes around the equilibrium state $X=x^\mathrm{eq}$, so
\begin{equation}
  \begin{split}
    d\mu^\mathrm{rel,M}(X)[\Gamma] &= d\mu^\mathrm{rel}(x^\mathrm{eq})[\Gamma] + \delta X \cdot \frac{\partial d\mu^\mathrm{rel}(x^\mathrm{eq})[\Gamma]}{\partial x^\mathrm{eq}} \\
    &= (1 - \delta \lambda \cdot \delta X(\Gamma) )\\
    &\quad \times \frac{\exp[ - \lambda(x^\mathrm{eq}) \cdot X(\Gamma)]}{Z(x^\mathrm{eq})} \, d\mu_L[\Gamma]\\
    &= (1 - \delta \lambda \cdot \delta X(\Gamma) ) \, d\mu^\mathrm{rel}(x^\mathrm{eq})[\Gamma]\\
    &= (1 - \delta \lambda (\Gamma) \cdot \delta X  ) \, d\mu^\mathrm{rel}(x^\mathrm{eq})[\Gamma],
  \end{split}
\end{equation}
where $\delta X(\Gamma) = X(\Gamma) - x^\mathrm{eq}$ and 
\begin{equation}
  \begin{split}
    \delta \lambda &= \delta X \cdot \frac{\partial \lambda(x^\mathrm{eq})}{\partial x^\mathrm{eq}}
      =  \delta X \cdot \left( \frac{\partial x^\mathrm{eq}}{\partial \lambda^\mathrm{eq}}\right)^{-1}\\
      &= -\left(\frac{\partial^2 \ln Z}{\partial \lambda^\mathrm{eq}\partial \lambda^\mathrm{eq}}\right)^{-1} \cdot  \delta X \\
    &= - \langle \delta X \, \delta X , \mu^\mathrm{rel}(x^\mathrm{eq})\rangle^{-1} \cdot \delta X,
  \end{split}
  \label{eq:expectation_X}
\end{equation}
such that
\begin{multline}
  d\mu^\mathrm{rel,M}(X)[\Gamma] = d\mu^\mathrm{rel}(x^\mathrm{eq})[\Gamma] \times \\ \biggl (1 + \delta X(\Gamma) \cdot \langle \delta X \, \delta X , \mu^\mathrm{rel}(x^\mathrm{eq})\rangle^{-1} \cdot \delta X \biggr ).
  \label{eq:Mori-measure}
\end{multline}
Because of the fact that $\delta X$ appears only linearly the Mori expectation value can be expressed as
\begin{equation}
  \begin{split}
    \langle A, \mu^\mathrm{rel,M}(X) \rangle &= A^\mathrm{eq} - \langle A \, \delta \lambda , \mu^\mathrm{rel}(x^\mathrm{eq}) \rangle \cdot \delta X\\
    &= A^\mathrm{eq} + \Omega_A \cdot \delta X,
  \end{split}
\end{equation}
with
\begin{equation}
  \Omega_A = \langle A \, \delta X , \mu^\mathrm{rel}(x^\mathrm{eq}) \rangle \cdot \langle \delta X \, \delta X , \mu^\mathrm{rel}(x^\mathrm{eq})\rangle^{-1}.
\end{equation}
Applying this operation multiple times gives the same result (since $\Omega_X=1$).
Therefore this operation defines a projection
\begin{equation}
  (\mathcal{P}^\mathrm{M} A)(X) = \langle A, \mu^\mathrm{rel,M}(X)\rangle.
\end{equation}
Due to the linearization around the equilibrium the result can be expected to be useful near equilibrium only.

For the Mori projection operator one always assumes that the equilibrium distribution is invariant, i.e., $\mathcal{L} \mu^\mathrm{rel}(x_\mathrm{eq}) = 0$.
This is obeyed if $\mathcal{L} \, (\lambda(x_\mathrm{eq}) \cdot X) =0$.
Therefore the linear combination given by $\lambda(x_\mathrm{eq}) \cdot X$ should be a conserved quantity.
When this requirement is obeyed one finds, using eq.~\eqref{eq:Mori-measure}, that
\begin{equation}
  \begin{split}
    \mathcal{P}^\mathrm{M} \mathcal{L} \, \tilde{A}^\mathrm{fluct}_{t-s} &= \langle \mathcal{L} \, \tilde{A}^\mathrm{fluct}_{t-s}, \mu^\mathrm{rel,M}(X_0) \rangle\\
    &= - \langle \tilde{A}^\mathrm{fluct}_{t-s}, \mathcal{L} \, \mu^\mathrm{rel,M}(X_0) \rangle \\
    &= \langle \tilde{A}^\mathrm{fluct}_{t-s} \dot{X}_0  , \mu^\mathrm{rel}(x^\mathrm{eq}) \rangle \cdot \\ &\quad \langle \delta X \, \delta X , \mu^\mathrm{rel}(x^\mathrm{eq})\rangle^{-1} \cdot \delta X,
  \end{split}
\end{equation}
where $\delta X = X_0 - x^\mathrm{eq}$.
Letting, $\exp[\mathcal{L} s]$ act on it, as in eq.~\eqref{eq:generalized_langevin}, gives $\delta X_s=\exp[\mathcal{L} s] (X_0 - x^\mathrm{eq})=X_s-x^\mathrm{eq}$. 
So, we have that
\begin{equation}
  \exp[\mathcal{L} s] \, \mathcal{P}^\mathrm{M} \mathcal{L} \, \tilde{A}^\mathrm{fluct}_{t-s} = \langle \tilde{A}^\mathrm{fluct}_{t-s} \dot{X}_0  , \mu^\mathrm{rel}(x^\mathrm{eq}) \rangle \cdot \lambda^\mathrm{M}(X_s).
  \label{eq:time-correlation_mori}
\end{equation}
The driving force can be seen as a linearization of $\lambda^\mathrm{M}(X)$ of the canonical $\lambda(X)$ around $x^\mathrm{eq}$ where it should be noted that the time correlation matrix times $\lambda^\mathrm{eq}$ equals zero because $\dot{X}_0\cdot\lambda^\mathrm{eq}=0$.
Alternatively one can introduce an entropy,
\begin{equation}
  \begin{split}
    S^\mathrm{M}(X) &= \lambda^\mathrm{eq} \cdot \delta X + \tfrac{1}{2} \langle \delta X \, \delta X , \mu^\mathrm{rel}(x^\mathrm{eq})\rangle^{-1} : (\delta X \, \delta X),\\
    \lambda^\mathrm{M}(X) &= \frac{\partial S^\mathrm{M}(X)}{\partial X}.
  \end{split}
  \label{eq:Mori-entropy}
\end{equation}
The final linear generalized Langevin equation, eq.~\eqref{eq:generalized_langevin}, that arises from the Mori-formalism is,
\begin{multline}
  A_t = A^\mathrm{eq} + \Omega_A \cdot \delta X_t + \\ \int_0^t \langle \tilde{A}^\mathrm{fluct}_{t-{t'}} \dot{X}_0  , \mu^\mathrm{rel}(x^\mathrm{eq}) \rangle \cdot \lambda^\mathrm{M}(X_{t'}) \, d{t'}+ \tilde{A}^\mathrm{fluct}_t,
  \label{eq:generalized_langevin_mori}
\end{multline}
The first two terms combined are the expectation value of $A(\Gamma)$ with respect to $\mu^\mathrm{rel,M}(X_t)$.
The last two term in eq.~\eqref{eq:generalized_langevin_mori} are clearly related by a fluctuation-dissipation relation.
The memory-integral term appears as a thermodynamic term.
Here the driving force is a derivative of the entropy.
The time-correlation appearing in the memory integral correlates fluctuations of $A$ with the time derivative of the macroscopic quantities $X$.

For the case of $A_t=X_t$ the equation becomes trivial.
Since $\tilde{X}^\mathrm{fluct}_0=X_0 - \mathcal{P} X_0=X_0-X_0=0$, $\tilde{X}^\mathrm{fluct}_t$ remains zero according to eq.~\eqref{eq:fluctuating_dynamics_solution}.
Also, $\Omega_X=1$, so eq.~\eqref{eq:generalized_langevin_mori} reduces to $X_t=X_t$.
The most commonly used equation for further approximations starts with $A_t= \mathcal{L} X_t$.
In this case one finds
\begin{equation}
  \dot{X}_t = \Omega \cdot \delta X_t + \\ \int_0^t \langle \dot{\tilde{X}}^\mathrm{fluct}_{t-{t'}} \dot{X}_0  , \mu^\mathrm{rel}(x^\mathrm{eq}) \rangle \cdot \lambda^\mathrm{M}(X_{t'}) \, d{t'} + \dot{\tilde{X}}^\mathrm{fluct}_t,
  \label{eq:mori_langevin}
\end{equation}
where $\Omega=\Omega_{\mathcal{L}X}$.

Further developments along these lines are due to Robertson \cite{Rob66} and Grabert \cite{Gra82}.
They introduced a linearization around a state $x_t$ (that evolves with time), so
\begin{equation}
  \begin{split}
    d\mu^\mathrm{rel,R}(X)[\Gamma] &\dot{=} d\mu^\mathrm{rel}(x_t + \delta X) [\Gamma] \\
    &= d\mu^\mathrm{rel}(x_t)[\Gamma] + \delta X \cdot \frac{\partial d\mu^\mathrm{rel}(x_t)[\Gamma]}{\partial x_t}, 
  \end{split}
  \label{eq:linearization_mu_rel}
\end{equation}
where $\dot{=}$ indicates that the subsequent expression should be linearized with respect to $\delta X$, and
\begin{multline}
  (\mathcal{P}^\mathrm{R}(t) A)(X) = \langle A, \mu^\mathrm{rel}(x_t)\rangle + \\
  \langle A \, \delta X , \mu^\mathrm{rel}(x_t) \rangle \cdot \langle \delta X \, \delta X , \mu^\mathrm{rel}(x_t)\rangle^{-1} \cdot \delta X,
  \label{eq:grabert_projection}
\end{multline}
with $\delta X=X-x_t$.
Operators like $\mathcal{L}$ and $\mathcal{P}(t)$ act on $X$, but not on $x_t$.
Here we used the assumed generalized canonical shape of $\mu^\mathrm{rel}$, eq.~\eqref{eq:canonical}.
Since $x_t$ is time dependent one finds a projection operator that is time-dependent.
Since (for all flavors of) $\mathcal{P}$, $\mathcal{P} \, X=X$, and the projection with $\mathcal{P}^\mathrm{R}$ results in a linear expression in $X$.

When constructing a fluctuating quantity the property one wants to satisfy is a generalization of eq.~\eqref{eq:projected_fluctuations}, namely,
\begin{equation}
  \mathcal{P}(0) \, \tilde{A}^\mathrm{fluct}_{0,t} = 0.
  \label{eq:projected_fluctuations_t}
\end{equation}
The reason one wants to use $\mathcal{P}(0)$ is because, as we will see further on, this corresponds to the initial ensemble.
The expectation values of fluctuations with respect to the initial ensembles are made to equal zero.
The generalization of the fluctuating term, eq.~\eqref{eq:fluctuating_dynamics_solution}, used by Grabert \cite{Gra82} that obeys this property is
\begin{equation}
  \tilde{A}^\mathrm{fluct}_{{t'},t} = \mathcal{Q}({t'}) \, \biggl\{ \mathcal{T}_{-}\exp \Bigl [\int_{t'}^t \mathcal{L} \mathcal{Q}({t''}) \, d{t''} \Bigr] \biggr \} \, A_0,
\end{equation}
so here $\mathcal{P}({t'}) \, \tilde{A}^\mathrm{fluct}_{{t'},t} =0$ for all ${t'}$. 
The exponent is reverse time-ordered.
This means that, by definition,
\begin{multline}
  \frac{d}{d{t'}} \mathcal{T}_{-}\exp \Bigl [\int_{t'}^t \mathcal{L} \mathcal{Q}({t''}) \, d{t''} \Bigr] = \\ - \mathcal{L}Q({t'}) \, \mathcal{T}_{-}\exp \Bigl [\int_{t'}^t \mathcal{L} \mathcal{Q}({t''}) \, d{t''} \Bigr].
\end{multline}
The decomposition as given in eq.~\eqref{eq:generalized_langevin} becomes
\begin{multline}
    A_t = \exp[\mathcal{L}t] \, \mathcal{P}(t) \, A_0 + \\ \int_0^t \, \exp[\mathcal{L} \, {t'}] \, \Bigl ( \mathcal{L} +\frac{d}{d{t'}} \Bigr) \, \tilde{A}^\mathrm{fluct}_{{t'},t} \, d{t'} + \tilde{A}^\mathrm{fluct}_{0,t} .
  \label{eq:decomposition_time_ordered}
\end{multline}
The differentiation to ${t'}$ gives an extra term (proportional to $\dot{\mathcal{Q}({t'})}$) besides the term $\mathcal{Q}({t'}) \mathcal{L} \tilde{A}^\mathrm{fluct}_{{t'},t}$.
To further simplify the resulting expression we need to use properties of the projection operator as introduced by Grabert.
This will be done below.

For the time-dependent projection-operator we find that
\begin{equation}
  \mathcal{P}^\mathrm{R}(t) \mathcal{P}^\mathrm{R}(t') = \mathcal{P}^\mathrm{R}(t').
\end{equation}
From this relation one can straightforwardly deduce that $\mathcal{Q}^\mathrm{R}(t) \mathcal{Q}^\mathrm{R}(t')=\mathcal{Q}^\mathrm{R}(t)$, and taking the derivative with respect to $t$ at $t'=t$ gives $\dot{\mathcal{Q}}^\mathrm{R}(t) \mathcal{Q}^\mathrm{R}(t)= \dot{\mathcal{Q}}^\mathrm{R}(t)$.

From the definition that $\mathcal{Q}^\mathrm{R}(t)$ is a projection operator one can find that $\dot{\mathcal{Q}}^\mathrm{R}(t) = \mathcal{P}^\mathrm{R}(t) \dot{\mathcal{Q}}^\mathrm{R}(t) \mathcal{Q}^\mathrm{R}(t) + \mathcal{Q}^\mathrm{R}(t) \dot{\mathcal{Q}}^\mathrm{R}(t) \mathcal{P}^\mathrm{R}(t)$.
Combining these two facts we have
\begin{equation}
  \dot{\mathcal{Q}}^\mathrm{R}(t) = \mathcal{P}^\mathrm{R}(t) \dot{\mathcal{Q}}^\mathrm{R}(t) \mathcal{Q}^\mathrm{R}(t).
\end{equation}
Using this relation to evaluate the derivative to $t'$ in eq.~\eqref{eq:decomposition_time_ordered} results into
\begin{equation}
  \begin{split}
    \Bigl ( \mathcal{L} +\frac{d}{d{t'}} \Bigr) \, \tilde{A}^\mathrm{fluct}_{{t'},t} &= ( \mathcal{L} - \mathcal{Q}^\mathrm{R}({t'}) \mathcal{L} + \dot{\mathcal{Q}}^\mathrm{R}({t'}) )  \, \tilde{A}^\mathrm{fluct}_{{t'},t}\\ &= \mathcal{P}({t'}) (\mathcal{L} + \dot{\mathcal{Q}}^\mathrm{R}({t'}) ) \tilde{A}^\mathrm{fluct}_{{t'},t}.
  \end{split}
\end{equation}
Inserting this equality and the definition of $\mathcal{P}^\mathrm{R}$ into eq.~\eqref{eq:decomposition_time_ordered} gives
\begin{multline}
  A_t \dot{=} \langle A, \mu^\mathrm{rel}(X_t) \rangle  + \\ \int_0^t \Bigl \langle(\mathcal{L} + \dot{\mathcal{Q}}^\mathrm{R}({t'}) ) \tilde{A}^\mathrm{fluct}_{{t'},t} , \mu^\mathrm{rel}(X_{t'}) \Bigr \rangle \, d{t'} + \tilde{A}^\mathrm{fluct}_{0,t}
  \label{eq:generalized_langevin_grabert}
\end{multline}
Here the linearization of $X_t$ should be taken around $x_t$ and $X_{t'}$ around $x_{t'}$.
The reason is that in eq.~\eqref{eq:decomposition_time_ordered} $\exp[\mathcal{L} {t'}] \mathcal{P}({t'})$ occur together.
The projection gives a linearization around $x_{t'}$, the operator $\exp[\mathcal{L} {t'}]$ transforms $X_0$ into $X_{t'}$.
The action of $\dot{\mathcal{Q}}^\mathrm{R}({t'})$ can be deduced from eq.~\eqref{eq:linearization_mu_rel} as,
\begin{equation}
  \dot{\mathcal{Q}}^\mathrm{R}({t'}) A = - \dot{\mathcal{P}}^\mathrm{R}({t'}) A = - \Bigl \langle A , \dot{x}_{t'} \cdot \frac{\partial^2\, \mu^\mathrm{rel}(x_{t'})}{\partial x_{t'} \partial x_{t'}} \Bigr \rangle \cdot (X-x_{t'}).
\end{equation}
Using this formula for performing the linearization of eq.~\eqref{eq:generalized_langevin_grabert} gives
\begin{equation}
  \begin{split}
    A_t &= \langle A, \mu^\mathrm{rel}(x_t) \rangle  + \int_0^t \langle \tilde{A}^\mathrm{fluct}_{{t'},t} \dot{X}_0 , \mu^\mathrm{rel}(x_{t'}) \rangle \cdot \lambda(x_{t'}) \, d{t'}\\
    &\quad + \Omega_A(x_t) \cdot \delta X_t\\
    &\quad - \int_0^t \Bigl \langle \tilde{A}^\mathrm{fluct}_{{t'},t} , \Bigl(\mathcal{L}^\dag +\dot{x}_{t'} \cdot \frac{\partial}{\partial x_{t'}}\Bigr) \frac{\partial \mu^\mathrm{rel}(x_{t'})}{\partial x_{t'}} \Bigr \rangle \cdot \delta X_{t'} \, d{t'}\\ &\quad + \tilde{A}^\mathrm{fluct}_{0,t}.
  \end{split}
    \label{eq:generalized_langevin_grabert_2}
\end{equation}
Here we still see the linearized character of the equation.
The full term $(\mathcal{L}^\dag +\dot{x}_{t'} \cdot {\partial}/{\partial x_{t'}})$ can be interpreted as a total time derivative.

Clearly one would like that $x_t$ closely follows $X_t$.
The usual choice, made by Robertson and Grabert, is to assume that $x_t$ equals the expectation value over an initial canonical distribution,
\begin{equation}
  x_t = \langle \exp[\mathcal{L}t] \, X, \mu^\mathrm{rel}(x_0) \rangle.
\end{equation}
With this choice $\langle \exp[\mathcal{L}t] \, \delta X_{t'}, \mu^\mathrm{rel}(x_0) \rangle=0$.
Using the same notation, also for quantities $a_t= \langle \exp[\mathcal{L}t] \, A, \mu^\mathrm{rel}(x_0) \rangle$, we find that
\begin{multline}
  a_t = \langle A, \mu^\mathrm{rel}(x_t) \rangle  + \\ \int_0^t \langle \mathcal{L} \tilde{A}^\mathrm{fluct}_{{t'},t} , \mu^\mathrm{rel}(x_{t'}) \rangle \, ds.
\end{multline}
Because all terms linear in $\delta X$ cancel.
An extra assumption used is to take $X_0=x_0$ then $\langle \tilde{A}^\mathrm{fluct}_{0,t}, \mu^\mathrm{rel}(x_0)\rangle=0$ as a consequence of requirement eq.~\eqref{eq:projected_fluctuations_t} and $\delta X_0=0$ inserted into eq.~\eqref{eq:grabert_projection}).
If this assumption is not made a term $\langle \tilde{A}^\mathrm{fluct}_t, \mu^\mathrm{rel}(x_0) \rangle$

For $\mu^\mathrm{rel}$ equal to the generalized canonical ensemble one has
\begin{equation}
  \begin{split}
    \langle \mathcal{L} \tilde{A}^\mathrm{fluct}_{s,t} , \mu^\mathrm{rel}(x_s) \rangle &= \langle \tilde{A}^\mathrm{fluct}_{s,t} , \mathcal{L} ^\dag \mu^\mathrm{rel}(x_s) \rangle\\
    &= \langle \tilde{A}^\mathrm{fluct}_{s,t} \dot{X}_0 , \mu^\mathrm{rel}(x_s) \rangle \cdot \lambda(x_s), 
  \end{split}
\end{equation}
such that
\begin{multline}
  a_t = \langle A, \mu^\mathrm{rel}(x_t) \rangle  + \int_0^t \langle \tilde{A}^\mathrm{fluct}_{{t'},t} \dot{X}_0 , \mu^\mathrm{rel}(x_{t'}) \rangle \cdot \lambda(x_{t'}) \, d{t'},
\end{multline}
when $X_0=x_0$.
The deviations from this average, given by $\delta A_t$, are similar (also linear) to the Mori generalized linear Langevin equation, eq.~\eqref{eq:mori_langevin}.
The difference is the linearization around a time-dependent state $x_t$.
This gives a few extra terms.

\acknowledgments{The author thanks H.~C.~\"{O}ttinger for feedback on an earlier version of the manuscript.}

\bibliography{../frank}

\end{document}